\documentclass[twocolumn,floatfix,showpacs,pra,amsmath,amssymb,superscriptaddress]{revtex4}

\usepackage{epsfig}
\input epsf.sty

\begin{document}

\markboth{S.E. Venegas-Andraca and S. Bose} {Quantum Walk-based
Generation of Entanglement Between Two Walkers}

\title{Quantum Walk-based Generation of Entanglement Between Two Walkers}

\author{Salvador E. Venegas-Andraca}
\address{Quantum Information Processing Group, Tecnol\'{o}gico de Monterrey Campus Estado de M\'{e}xico,
Carretera Lago Gpe. Km 3.5, Atizap\'{a}n de Zaragoza, Edo. M\'{e}xico, 52926, M\'{e}xico\\
sva@mindsofmexico.org, salvador.venegas-andraca@keble.oxon.org}

\author{Sougato Bose}
\address{Department of Physics and Astronomy, University College
of London. Gower Street, London WC1E 6BT, United Kingdom\\
sougato@theory.phys.ucl.ac.uk}

\begin{abstract}

Quantum walks can be used either as tools for quantum algorithm
development or as entanglement generators, potentially useful to
test quantum hardware. We present a novel algorithm based on a
discrete Hadamard quantum walk on a line with one coin and two
walkers whose purpose is to generate entanglement between walkers.
We provide several classical computer simulations of our quantum
algorithm in which we show that, although the asymptotical amount
of entanglement generated between walkers does not reach the
highest degree of entanglement possible at each step for either
coin measurement outcome, the entanglement ratio (entanglement
generated/highest value of entanglement possible, for each step)
tends to converge, and the actual convergence value depends on the
coin initial state and on the coin measurement outcome.
Furthermore, our numerical simulations show that, for the quantum
walks used in our algorithm, the value towards which entanglement
ratio converges also depends on the position probability
distribution symmetry of a quantum walk computed with one single
walker and the same coin initial state employed in the
corresponding quantum walk with two walkers.
\end{abstract}

\maketitle

\section{Introduction}

Quantum walks were designed as quantum counterparts of classical random walks,
a branch of stochastic processes widely used in algorithm development. Although some authors
have selected the name \lq \lq quantum random walk'' to refer to quantum phenomena \cite{godoy92,gudder88,konno02a}
and, in fact, in the seminal work
by R.P. Feynman \cite{feynman86} about quantum mechanical computers we find a proposal
that could be interpreted as a (continuous-time) quantum walk \cite{chase08}, it is generally
accepted that the first paper with quantum walks as its main topic was published
in 1993 by Aharonov {\it et al} \cite{aharonov93}. Thus, the links between
classical random walks and quantum walks, as well as the utility of quantum walks
in computer science, are two fresh and open areas of research.

Since one of the main goals in quantum computing is the
development of quantum algorithms, and given the success of
employing classical random walks for computing solutions to
NP-complete problems \cite{schoning99,iwama03,motwani95}, there
has been a huge interest in understanding the physical and
computational properties of quantum walks over the last few years
on both experimental
\cite{du03,sanders03,agarwal05,chandrashekar06,rai08} and
theoretical research communities (see \cite{sva_book_qw}
for a review on theoretical aspects of quantum walks). In addition
to their usage in computer science, the study of quantum walks is
relevant to the modelling of physical phenomena such as energy
transfer in photosynthetic systems \cite{mohseni08}. Moreover,
although it has been proved that certain properties of quantum
walks are also reproducible by classical systems (like variance
enhancement with respect to classical random walks
\cite{knight03,knight03b,knight04}), it is also true that uniquely
quantum mechanical properties of quantum walks, such as
entanglement, may be employed to building methods in order to test
the \lq \lq quantumness'' of emerging technologies for the
creation of quantum computers. Thus it is of crucial importance to
develop methods of entanglement generation through quantum walks
so that the genuine quantum nature of a given walk with given
physical systems may be tested.

Quantum entanglement has been incorporated into quantum walks research either as
a result of performing a quantum walk \cite{carneiro05,maloyer07,abal05,abal07,goyal09,annabestani09}
or as a resource to build new kinds of quantum walks \cite{paunkovic04,sva05,chandrashekar06b,liu08}. Since entanglement
is a key component in quantum computation, it is worth keeping in mind that quantum walks
can be used either as entanglement generators or as computational processes taking advantage
of this quantum mechanical property.


In this paper we introduce a novel algorithm based on a discrete
quantum walk on a line with one coin and two walkers whose purpose
is to generate entanglement between walkers. After evolving the
quantum walk for a certain number of steps, we perform a
measurement on the coin state. We then obtain a post-measurement
quantum state composed by the tensor product of one coin state and
several walker components. We take the walker components of this
post-measurement state and calculate the entanglement between
walkers. We perform many quantum walks with the same initial
conditions and evolution operators, so that we have a quantum walk
ready to be measured for each time step. In addition to our
algorithm, we provide several simulation results using different
initial conditions for the proposed quantum walks. While this
analysis highlights the potential of a quantum walk to entangle
high dimensional quantum systems (the dimension of the space
available to the walkers grow in each step) it can also at times
be practically useful. This will be the case when the walkers are
systems which do not directly interact with each other such as two
different electromagnetic field modes. Then the coin can be a
common system such as an atom which interact with both and can
entangle them to a high degree with the degree depending on the
number of steps possible within the reasonable decoherence time of
the fields.

\section{Algorithm for entanglement generation }

In this section we present our algorithm for the generation
of entanglement in a family of quantum walks on an unrestricted
line. A succint mathematical representation of a quantum walk
after $n$ steps is

\begin{equation}
|\psi \rangle_n = ({\hat U})^n |\psi\rangle_\text{initial},
\label{qw_definition}
\end{equation}

where $|\psi\rangle_{\text{initial}}$ is the initial total
state of the quantum walk. In our case, the family of quantum walks
we shall employ is composed by the tensor product of
one coin and two walkers

\begin{equation}
|\text{coin}\rangle \otimes |\text{walker}_1,\text{walker}_2\rangle
\end{equation}

as total initial state.
After several applications of an evolution operator composed of
a coin operator and a shift operator, we perform a measurement
on the coin state. The result of this operation is a post-measurement
quantum state composed by the tensor product of one coin state and
several walker components. We take the walker components of this
coin post-measurement state and calculate the entanglement between walkers
using the von Neumann entropy

\begin{equation}
E(|\psi\rangle) = S(\rho_A) = S(\rho_B) = - \sum_{i=1}^d \alpha_i^2 \log_2 (\alpha_i^2).
\label{von_neumann}
\end{equation}

where $|\psi \rangle = \sum_{i=1}^d \alpha_i |i_A\rangle |i_B\rangle$ is the Schmidt decomposition
of a bipartite quantum state $|\psi \rangle$.
We compute $n$ quantum walks using the same initial states and evolution
operator in order to measure the degree of entanglement between walkers
{\it for each step}, so that the final result of this algorithm is
a graph with the amount of entanglement available at each step.
We summarize this explanation in algorithm 1.
\\
\\
{\bf Algorithm 1. Quantification of entanglement}.
\\
Input: A maximum number of steps $n$ for the quantum walk, and $n$ identically prepared total initial states $|\psi\rangle_0$ with one coin and two walkers.
\\
Objective: To quantify the amount of entanglement between walkers for each step of the quantum walk.
\\
01.  Set t=1
\\
02.  While ($t \leq n$)
\\
03. Apply the evolution operator $\hat{U}^t = (\hat{S}(\hat{C} \otimes \hat{I}))^t$ to $|\psi\rangle_0$.
\\
04. Perform a measurement on the coin system.
    Since $|\text{coin}\rangle \in {\cal H}^2$ there are only two possible outcomes. We label them $\alpha_0$ and $\alpha_1$.
\\
05. For outcome $\alpha_0$ then
\\
06. Compute the post-measurement quantum state $|\psi\rangle_{t,pm}^{c_0}$
\\
07. Quantify entanglement between walkers from quantum state $|\psi\rangle_{t,pm}^{c_0}$
\\
08. For outcome $\alpha_1$ then
\\
09. Compute the post-measurement quantum state $|\psi\rangle_{t,pm}^{c_1}$
\\
10. Quantify entanglement between walkers from quantum state $|\psi\rangle_{t,pm}^{c_1}$
\\
11. Increase t by 1
\\

As stated in the introduction, we are interested in quantifying the amount
of entanglement between walkers for each coin outcome, as well as in studying
the impact of different initial quantum states in this quantification of entanglement.
The following lines shows corresponding results using unrestricted quantum walks
on a line.

\subsection{Entanglement Generation in unrestricted Quantum Walks on a Line}

We shall use Eqs. (\ref{zero_initial_condition})-(\ref{five_initial_condition})  as total initial states,
where each initial condition has the form
$|\psi\rangle_0 = |\text{coin}\rangle_0 \otimes |\text{position}\rangle_0$,
with $|\text{coin}\rangle_0$ as coin initial state and $|\text{position}\rangle_0$
as walker initial state.

\begin{subequations}
\begin{equation}
|\psi\rangle_0 =|0\rangle_c\otimes |0,0\rangle_p
\label{zero_initial_condition}
\end{equation}
\begin{equation}
|\psi\rangle_0 =  |1\rangle_c\otimes |0,0\rangle_p
\label{two_initial_condition}
\end{equation}
\begin{equation}
|\psi\rangle_0 = ({1 \over \sqrt{2}}|0\rangle_c + {i \over \sqrt{2}}|1\rangle_c) \otimes |0,0\rangle_p
\label{three_initial_condition}
\end{equation}
\begin{equation}
|\psi\rangle_0 = ({i \over \sqrt{2}}|0\rangle_c + {1 \over \sqrt{2}}|1\rangle_c) \otimes |0,0\rangle_p
\label{fourth_initial_condition}
\end{equation}
\begin{equation}
|\psi\rangle_0 = (\sqrt{0.85}|0\rangle_c - \sqrt{0.15}|1\rangle_c) \otimes |0,0\rangle_p
\label{five_initial_condition}
\end{equation}
\end{subequations}


where subindex $c$ stands for \lq coin' and  subindex $p$ stands for \lq walker position'.
\\

Additionally, we use the Hadamard operator as coin operator

\begin{equation}
{\hat H}= {1 \over \sqrt{2}}(|0\rangle_c \langle 0| + |0\rangle_c \langle 1| + |1\rangle_c \langle 0| -|1\rangle_c \langle 1|)
\label{hadamard_chapter_qwec_ii}
\end{equation}

Our shift operator is given by

\begin{multline}
{\hat S}_{\text{ent}}= |0\rangle_c \langle 0| \otimes \sum_i |i+1, i+1\rangle_p \langle i, i| + \\
|1\rangle_c \langle 1| \otimes \sum_i |i-1, i-1\rangle_p \langle i, i|
\label{shift_operator_quant_entanglement}
\end{multline}

The observable used for coin measurement (step 4 of algorithm 1) is given by

\begin{equation}
{\hat M} = \alpha_0{\hat M}_0 + \alpha_1{\hat M}_1 = \alpha_0|0\rangle_c \langle 0| + \alpha_1|1\rangle_c \langle 1|
\label{observable_qw}
\end{equation}

With the purpose of exemplifying the behavior of algorithm 1,
we show in the following lines three steps of a quantum walk and
corresponding entanglement measurement using Eq. (\ref{zero_initial_condition})
as total initial state, and Eqs. (\ref{hadamard_chapter_qwec_ii}) and
(\ref{shift_operator_quant_entanglement})
as corresponding coin and shift operators.
Using Eq. (\ref{qw_definition}) we find that

\begin{equation}
|\psi\rangle_1 = {1 \over \sqrt{2}} (|0\rangle_c |1,1\rangle_p + |1\rangle_c |-1,-1\rangle_p)
\label{state_psi_1}
\end{equation}

\begin{equation}
|\psi\rangle_2 = {1 \over 2} (|0\rangle_c |2,2\rangle_p + |1\rangle_c |0,0\rangle_p +
                              |0\rangle_c |0,0\rangle_p - |1\rangle_c |-2,-2\rangle_p)
\label{state_psi_2}
\end{equation}


\begin{multline}
|\psi\rangle_3 =
{1 \over 2\sqrt{2}}(|0\rangle_c |3,3\rangle_p   + |1\rangle_c |1,1\rangle_p + |0\rangle_c |1,1\rangle_p \\
                  - |1\rangle_c |-1,-1\rangle_p + |0\rangle_c |1,1\rangle_p + |1\rangle_c |-1,-1\rangle_p \\
                  - |0\rangle_c |-1,-1\rangle_p + |1\rangle_c |-3,-3\rangle_p)
\label{state_psi_3}
\end{multline}

For $|\psi\rangle_1$ (Eq. (\ref{state_psi_1})), the post-measurement quantum state
after performing a coin measurement with measurement operator ${\hat M}_0$
(Eq. (\ref{observable_qw})) is given by $|\psi\rangle_{1,pm}^{c_0}=|0\rangle_c |1,1\rangle_p$,
and the degree of entanglement between walkers is clearly $0$. As for coin $1$, we perform a
coin measurement on $|\psi\rangle_1$ (Eq. (\ref{state_psi_1})) using measurement operator
${\hat M}_1$ (Eq. (\ref{observable_qw})), obtaining as post-measurement quantum state
$|\psi\rangle_{1,pm}^{c_1}=|1\rangle_c |-1,-1\rangle_p$. It is also clear that the degree
of entanglement between walkers in $|\psi\rangle_{1,pm}^{c_1}$ is $0$.

In step 2 (Eq. (\ref{state_psi_2})), we have $|\psi\rangle_{2,pm}^{c_0} = {1 \over \sqrt{2}} |0\rangle_c(|2,2\rangle_p + |0,0\rangle_p) $
as coin $|0\rangle_c$ post-measurement state, and corresponding entanglement betwen walkers
is equal to $1$, since ${1 \over \sqrt{2}}(|2,2\rangle_p + |0,0\rangle_p)$ is a maximally entangled state.
Along the same lines, the coin $|1\rangle_c$ post-measurement state is given by
$|\psi\rangle_{2,pm}^{c_1} = {1 \over \sqrt{2}} (|1\rangle_c)(|0,0\rangle_p + |-2,-2\rangle_p)$.
Since ${1 \over \sqrt{2}}(|0,0\rangle_p + |-2,-2\rangle_p)$ is a maximally entangled state,
its degree of entanglement is equal to $1$.

Finally, in step 3 (Eq. (\ref{state_psi_3})),
$|\psi\rangle_{3,pm}^{c_0} = {1 \over \sqrt{6}}|0\rangle_c(|3,3\rangle_p + 2|1,1\rangle_p - |-1,-1\rangle_p)$,
and corresponding degree of entanglement between walkers is equal to $1.2516$ (maximum degree of
entanglement attainable between walkers is $\log_2 3 = 1.585$.) As for coin $|1\rangle_c$,
$|\psi\rangle_{3,pm}^{c_1} = {1 \over \sqrt{2}}(|0\rangle_c)(|1,1\rangle_p + |-3,-3\rangle_p)$,
with degree of entanglement between walkers equal to $1$.

\begin{figure}
(i)\epsfig{width=1.5in,file=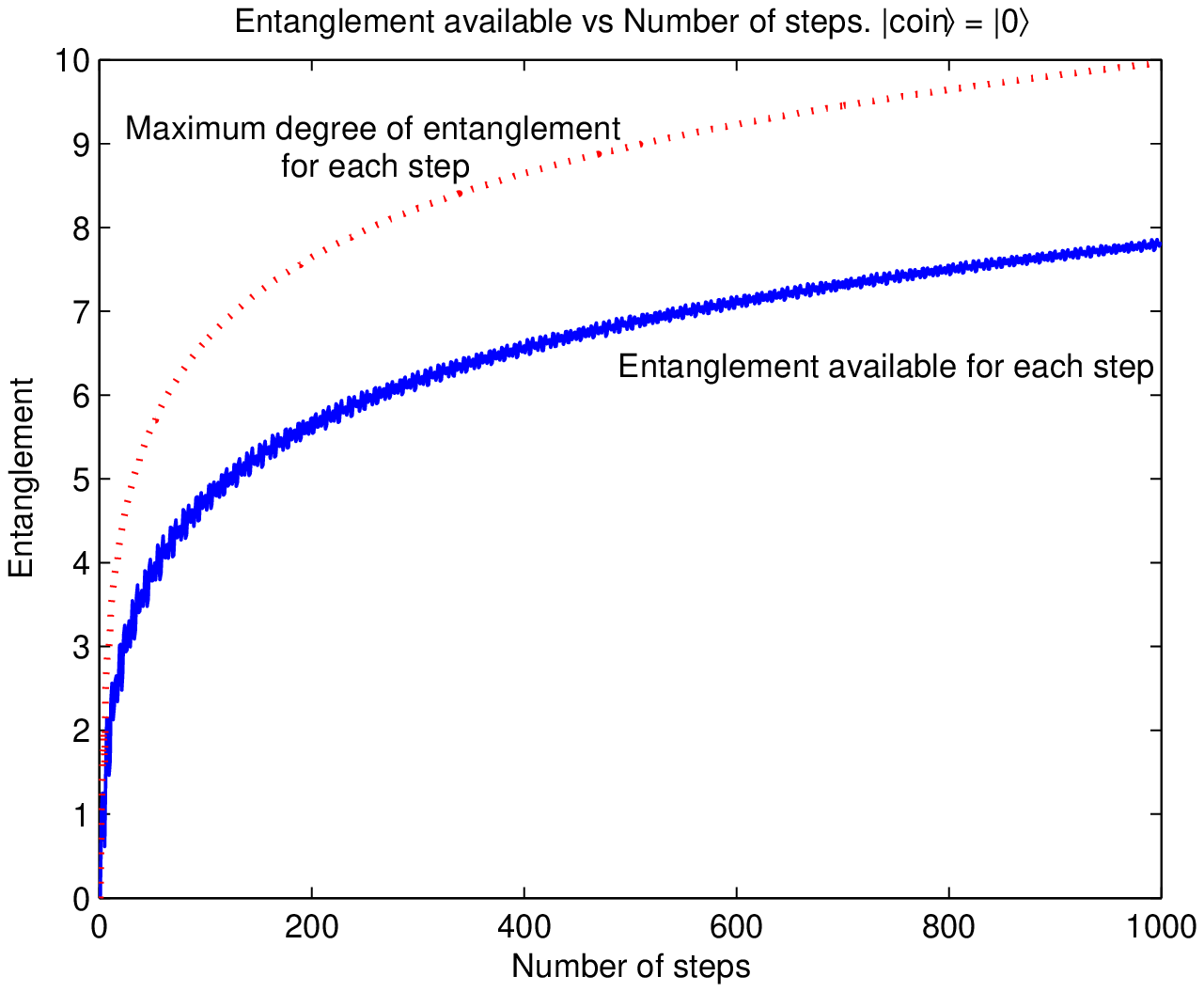}
(ii)\epsfig{width=1.5in,file=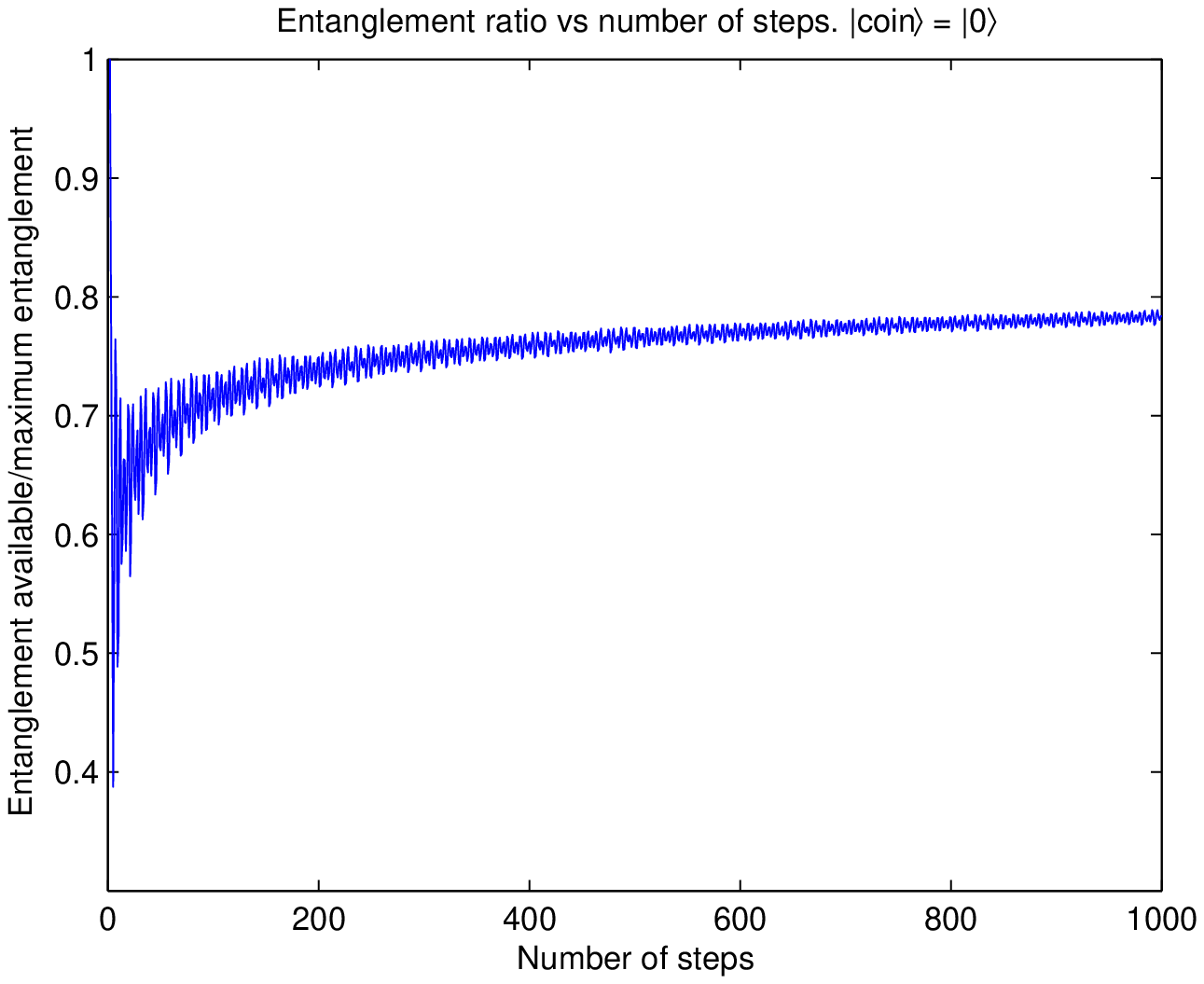}
\caption{{\small After computing a  1000-steps quantum walk
$|\psi\rangle_{1000} = [{\hat S_{\text{ent}}} (\hat{H} \otimes {\hat I} )]^{1000} |\psi\rangle_0$
with $|\psi\rangle_0$ given by Eq. (\ref{zero_initial_condition}) and
Eqs. (\ref{hadamard_chapter_qwec_ii}) and
(\ref{shift_operator_quant_entanglement})
as coin (${\hat H}$) and shift (${\hat S}$) operators,
we perform a coin measurement on $|\psi\rangle_{1000}$
using measurement operator ${\hat M}_0$ (Eq. (\ref{observable_qw})).
The thin line of (i) (red color online) shows the maximum
degree of entanglement between walkers attainable in the post-measurement
quantum state $|\psi\rangle_{t,pm}^{c_0}$ (for example, $\log_2 2 = 1$
for the second step and $\log_2 3 = 1.585$ for the third step), and
the thick line of (i) (blue color online) shows the actual entanglement between walkers available
at each step. We can see that, asymptotically, the entanglement available
is about 80\% of the corresponding maximum degree of entanglement (plot (ii)).
}}\label{first_condition_coin_zero}
\end{figure}

\begin{figure}[hbt]
(i)\epsfig{width=1.5in,file=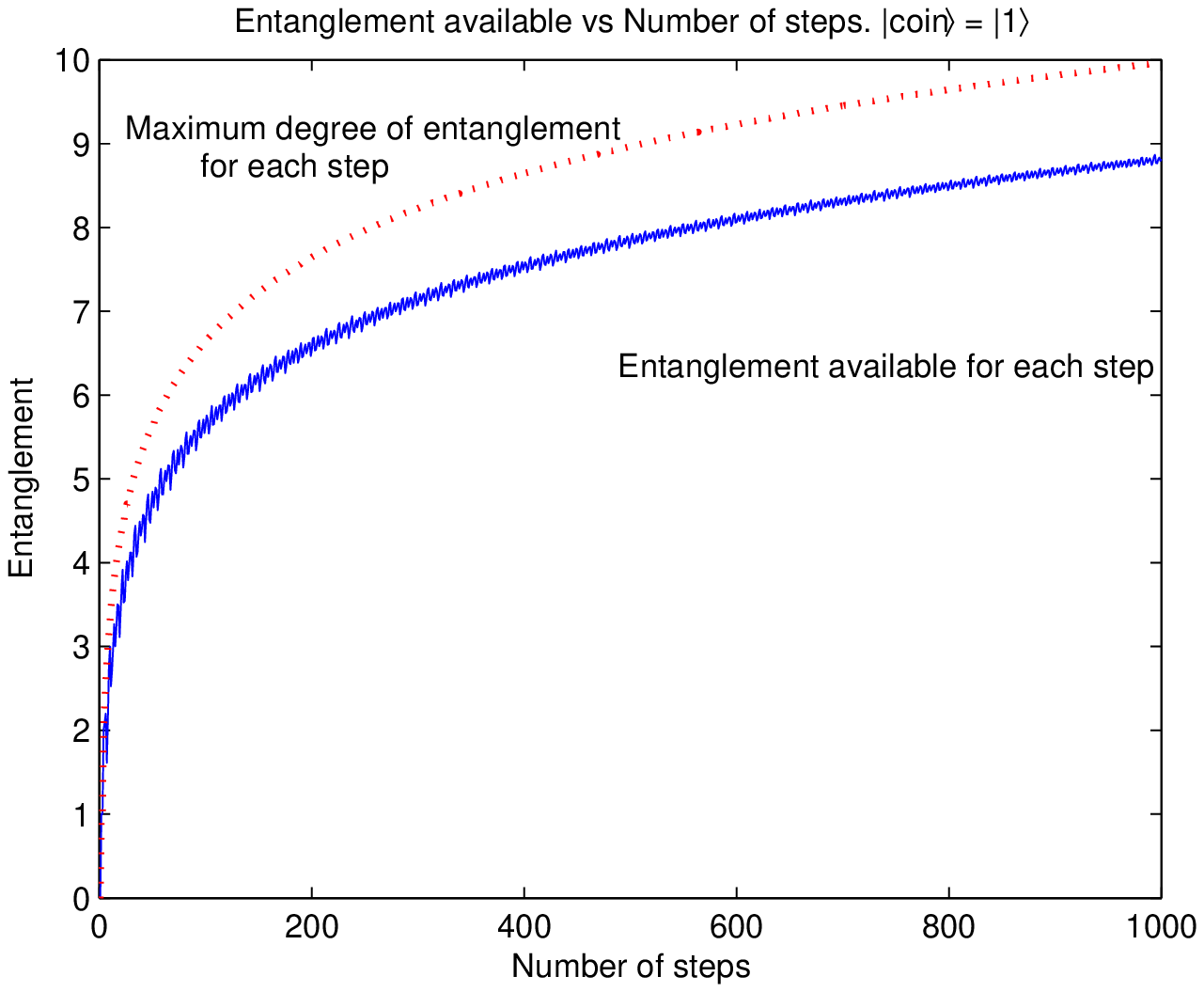}
(ii)\epsfig{width=1.5in,file=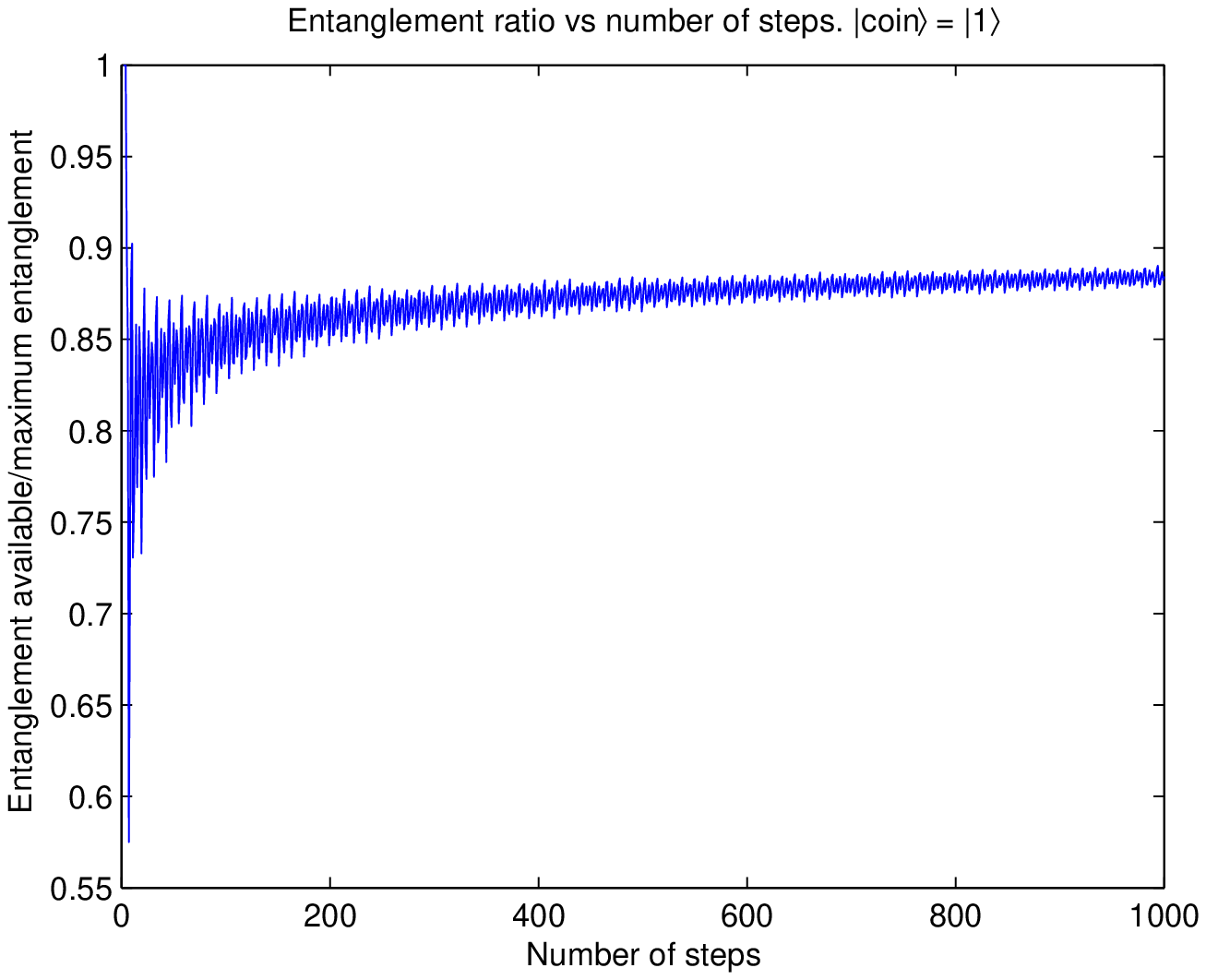}
\caption{{\small After computing of a 1000-steps
quantum walk $|\psi\rangle_{1000} = [{\hat S_{\text{ent}}} (\hat{H} \otimes {\hat I} )]^{1000} |\psi\rangle_0$
with $|\psi\rangle_0$ given by Eq. (\ref{zero_initial_condition}) and
Eqs. (\ref{hadamard_chapter_qwec_ii}) and
(\ref{shift_operator_quant_entanglement})
as coin (${\hat H}$) and shift (${\hat S}$) operators,
we perform a coin measurement on $|\psi\rangle_{1000}$
using measurement operator ${\hat M}_1$ (Eq. (\ref{observable_qw})).
The thin curve of (i) (red color online) shows the maximum
degree of entanglement between walkers attainable in the post-measurement
quantum state $|\psi\rangle_{t,pm}^{c_1}$ (for example, $\log_2 2 = 1$
for the second step and $\log_2 3 = 1.585$ for the third step), and
the thick curve of (i) (blue color online) shows the actual entanglement between walkers available
at each step. We can see that, for large number of steps, the entanglement available
is about 90\% of the corresponding maximum degree of entanglement (graph (ii)).
}}
\label{first_condition_coin_one}
\end{figure}

\begin{figure}[hbt]
(i)\epsfig{width=1.5in,file=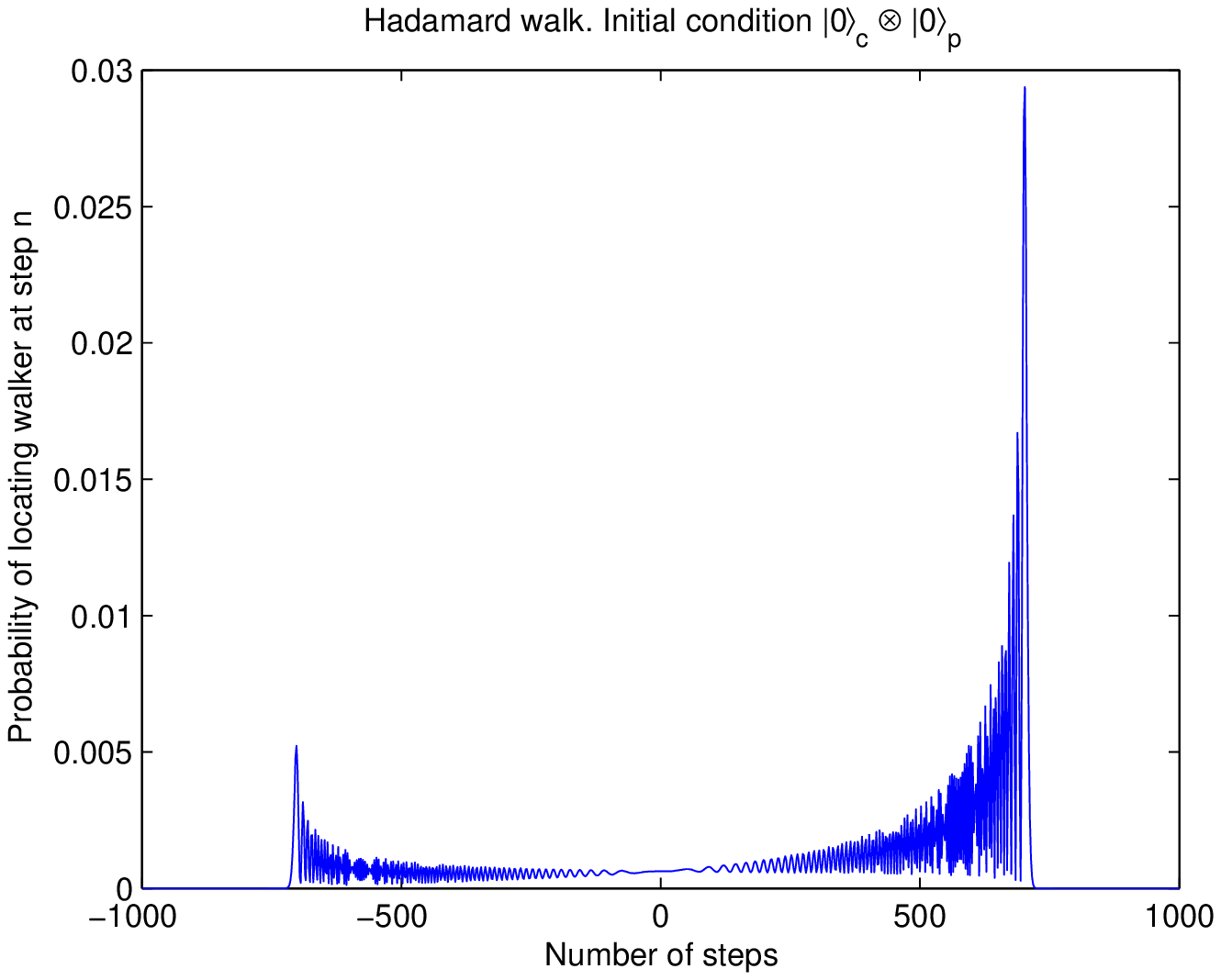}
(ii)\epsfig{width=1.5in,file=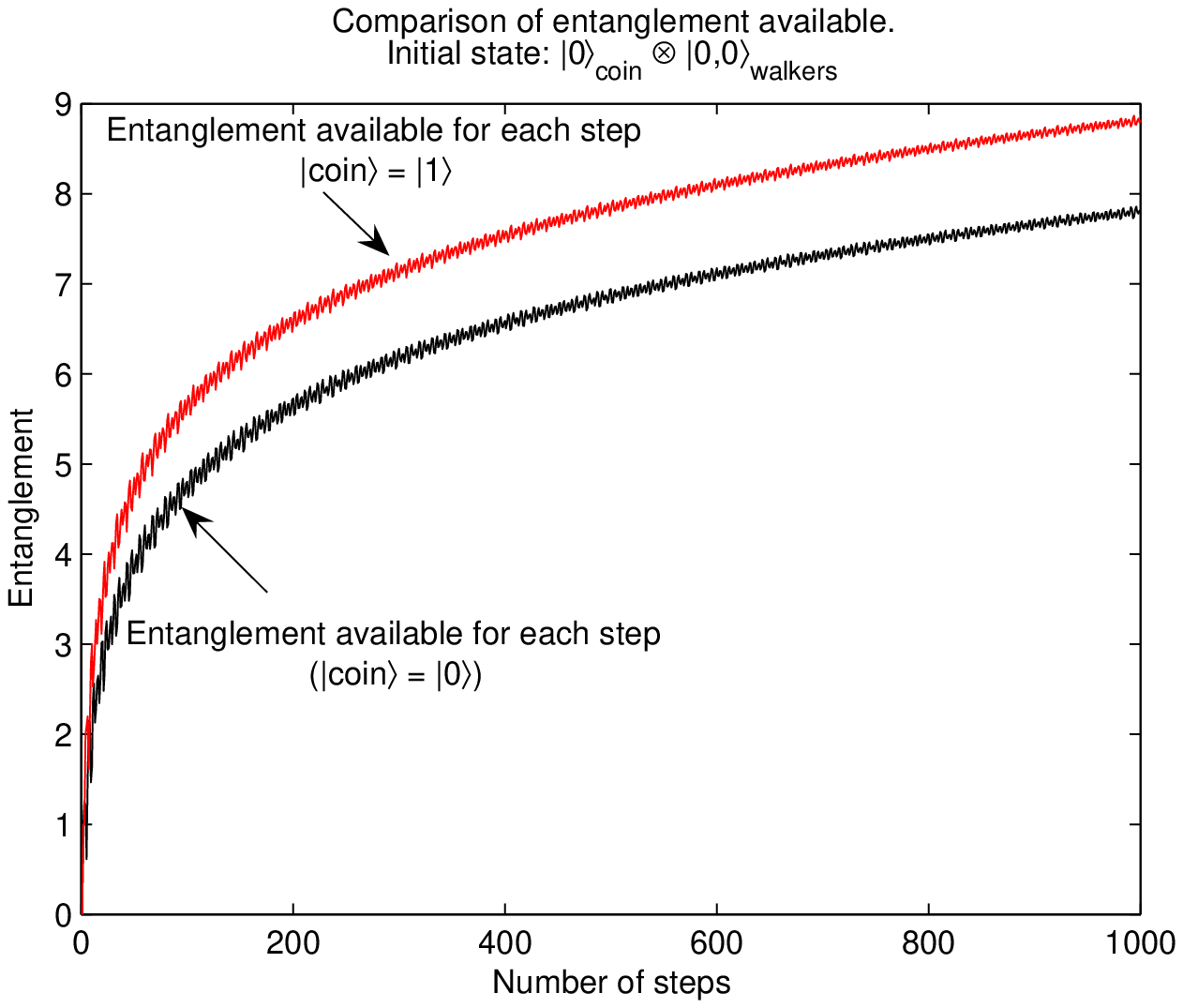}
\caption{{\small Plot (i) presents the probability vs location graph of
a 1000-step Hadamard quantum walk with an initial state
$|0\rangle_c \otimes |0\rangle_p$ and shift operator provided
by Eq. (\ref{shift_single}). The symmetry of this walk, about a line
passing through the origin and perpedicular to the $x$ axis, is the same
as that of a Hadamard quantum walk with initial state given by
$|\psi\rangle = |0\rangle_c \otimes |0,0\rangle_p$ and
shift operator given by Eq. (\ref{shift_operator_quant_entanglement}).
Plot (ii) is a summary of Figs. (\ref{first_condition_coin_zero}.i)
and (\ref{first_condition_coin_one}.i), and shows that the amount
of entanglement between walkers available in post-measurement state
$|\psi\rangle_{t,pm}^{c_1}$ tends to be higher than the amount of entanglement
between walkers available in post-measurement state $|\psi\rangle_{t,pm}^{c_0}$.
}}\label{first_condition_entanglement_comparison}
\end{figure}

We show in Figs. (\ref{first_condition_coin_zero}), (\ref{first_condition_coin_one})
and (\ref{first_condition_entanglement_comparison}),
simulation results for a $1000$-steps quantum walk performed with
Eq. (\ref{zero_initial_condition}) as total initial state and
Eqs. (\ref{hadamard_chapter_qwec_ii}) and (\ref{shift_operator_quant_entanglement})
as coin and shift operators.

Fig. (\ref{first_condition_coin_zero}) presents the results of
measuring entanglement between walkers in a coin $|0\rangle_c$
post-measurement state $|\psi\rangle_{t,pm}^{c_0}$.
In Fig. (\ref{first_condition_coin_zero}.i) we show two curves.
The thin curve (red color online) indicates, for each step of the quantum walk,
the maximum amount of entanglement between walkers achievable
at each time step, while the thick curve (blue color online) shows the actual degree
of entanglement between walkers available for each step. We can
see that, as the number of steps increases, the amount of entanglement
available vs the maximum degree of entanglement attainable is about
80\% (Figure (\ref{first_condition_coin_zero}. ii).)

In Fig. (\ref{first_condition_coin_one}) we present the same results
as in Fig. (\ref{first_condition_coin_zero}) but for a coin $|1\rangle_c$
post-measurement state $|\psi\rangle_{t,pm}^{c_1}$. First of all, we notice
that, as in the previous paragraph, the degree of entanglement between walkers
available in $|\psi\rangle_{t,pm}^{c_1}$ (thin line (red color online) of Fig. (\ref{first_condition_coin_one}.i))
does not reach the highest degree of entanglement attainable at each time step
(thick (blue color online) line in Fig. (\ref{first_condition_coin_one}.i)). However, it can be seen by
comparing the asymptotical behavior shown in Fig. (\ref{first_condition_coin_zero}.i)
and Fig. (\ref{first_condition_coin_one}.i) that, if the coin measurement
outcome is $\alpha_1$ (Fig. (\ref{first_condition_coin_one}.i)) then
the amount of entanglement available between walkers tends to be higher
(about 90\%, Fig. (\ref{first_condition_coin_one}.ii))
than the corresponding degree of entanglement between walkers
for a coin measurement outcome $\alpha_0$
(Fig. (\ref{first_condition_coin_zero}.i)) which is, as shown
in Fig. (\ref{first_condition_coin_zero}.ii), about 80\%.

In Fig. (\ref{first_condition_entanglement_comparison}.i) we display
the probability vs location graph of a 1000-step Hadamard
quantum walk
with an initial state given by $|0\rangle_c \otimes |0\rangle_p$ (i.e. one coin and only one walker)
and shift operator provided by

\begin{equation}
{\hat S} = |0\rangle_c \langle 0| \otimes \sum_i |i+1 \rangle_p \langle i|
+ |1\rangle_c \langle 1| \otimes \sum_i |i-1 \rangle_p \langle i|.
\label{shift_single}
\end{equation}

The symmetry of this walk, about a line passing through the origin and perpendicular
to the $x$ axis, is the same as that of a Hadamard quantum walk with
initial state given by $|\psi\rangle = |0\rangle_c \otimes |0,0\rangle_p$
and shift operator given by Eq. (\ref{shift_operator_quant_entanglement}).
The black curve of Fig. (\ref{first_condition_entanglement_comparison}.ii)
shows the amount of entanglement available between walkers in the post-measurement
state $|\psi\rangle_{t,pm}^{c_0}$ (as in Fig. (\ref{first_condition_coin_zero}.i)),
while the gray curve (red color online) shows the corresponding degree of entanglement available
between walkers for post-measurement state $|\psi\rangle_{t,pm}^{c_1}$
(as in Fig. (\ref{first_condition_coin_one}.i)). The purpose of
Fig. (\ref{first_condition_entanglement_comparison}) is to relate the amount
of entanglement available for each coin post-measurement state with the
symmetry of the quantum walk and, consequently, with the total initial
state of the quantum walk. We shall come back to
Fig. (\ref{first_condition_entanglement_comparison}) shortly.

\begin{figure}[hbt]
(i)\epsfig{width=1.5in,file=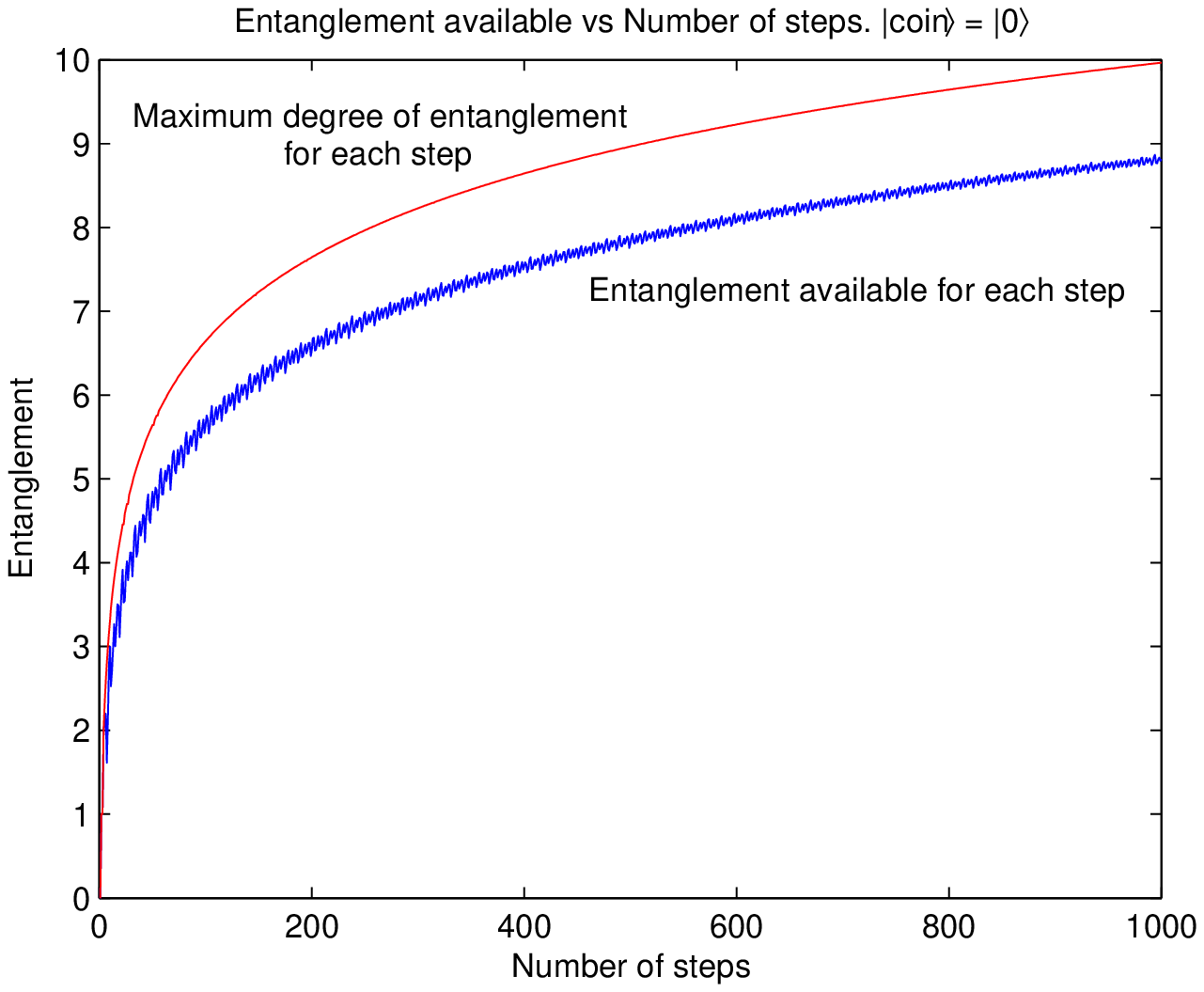}
(ii)\epsfig{width=1.5in,file=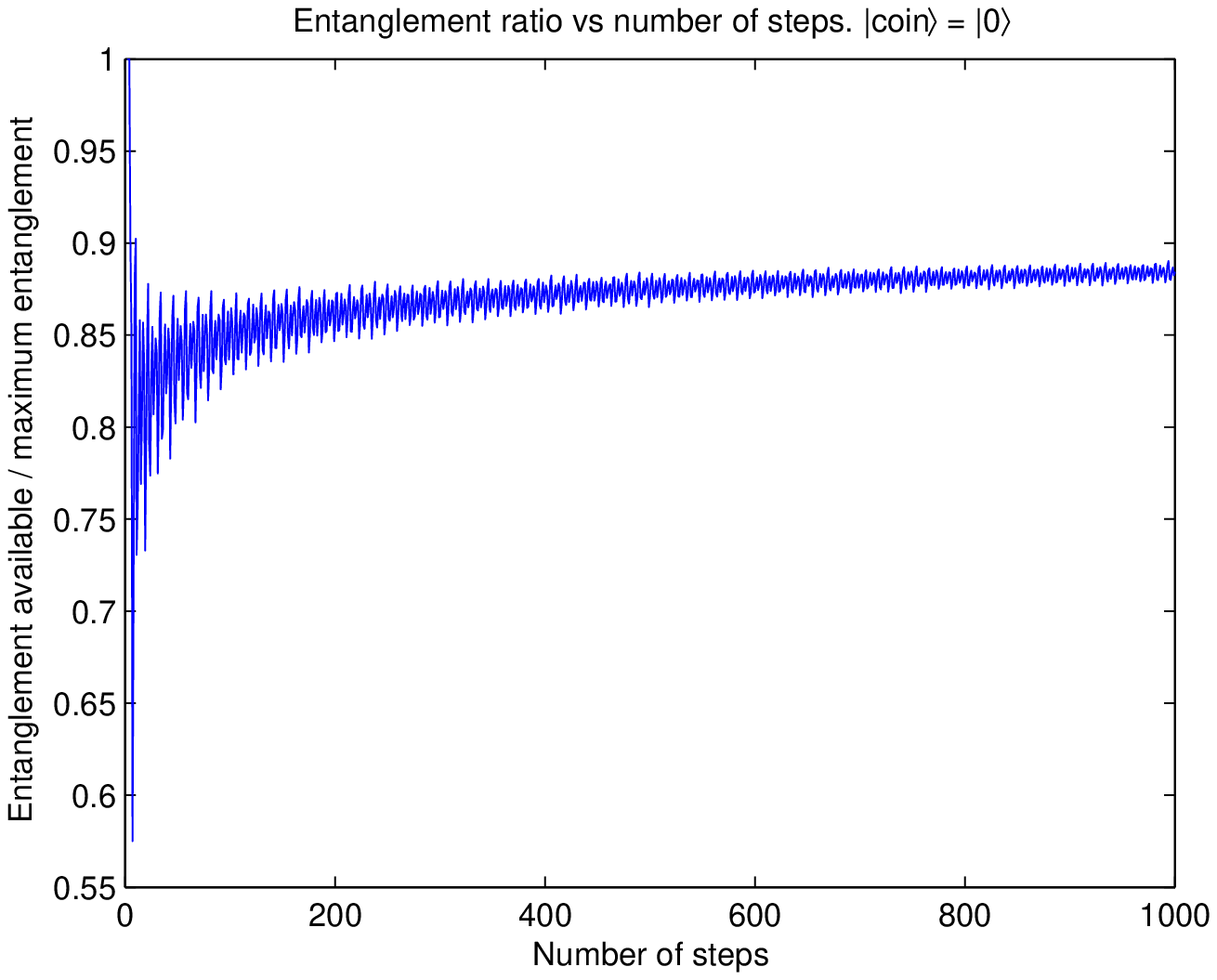}
\caption{{\small Entanglement values for coin post-measurement state
$|\psi\rangle_{t,pm}^{c_0}$ computed from a 1000-steps quantum walk
$|\psi\rangle_{1000} = [{\hat S_{\text{ent}}} (\hat{H} \otimes {\hat I} )]^{1000} |\psi\rangle_0$
with $|\psi\rangle_0$ given by Eq. (\ref{two_initial_condition}),
Eqs. (\ref{hadamard_chapter_qwec_ii}) and (\ref{shift_operator_quant_entanglement})
as coin (${\hat H}$) and shift (${\hat S}$) operators,
and measurement operator ${\hat M}_0$ (Eq. (\ref{observable_qw})).
The thin line of (i) (red color online) shows the maximum degree of entanglement between
walkers attainable in the post-measurement quantum state $|\psi\rangle_{t,pm}^{c_0}$,
and the thick line of (i) (blue color online) shows the actual entanglement between walkers available
at each step. We can see that, asymptotically, the entanglement available
is about 90\% of the corresponding maximum degree of entanglement (plot (ii)).
Note that this amount of entanglement available between walkers (90\%) is {\it higher}
than the amount of entanglement available between walkers (80\%) for coin $|0\rangle_c$
post-measurement quantum state with initial state $|0\rangle_c \otimes |0,0\rangle_p$
(Fig. (\ref{first_condition_coin_zero})).
}}
\label{second_condition_coin_zero}
\end{figure}

\begin{figure}[hbt]
(i)\epsfig{width=1.5in,file=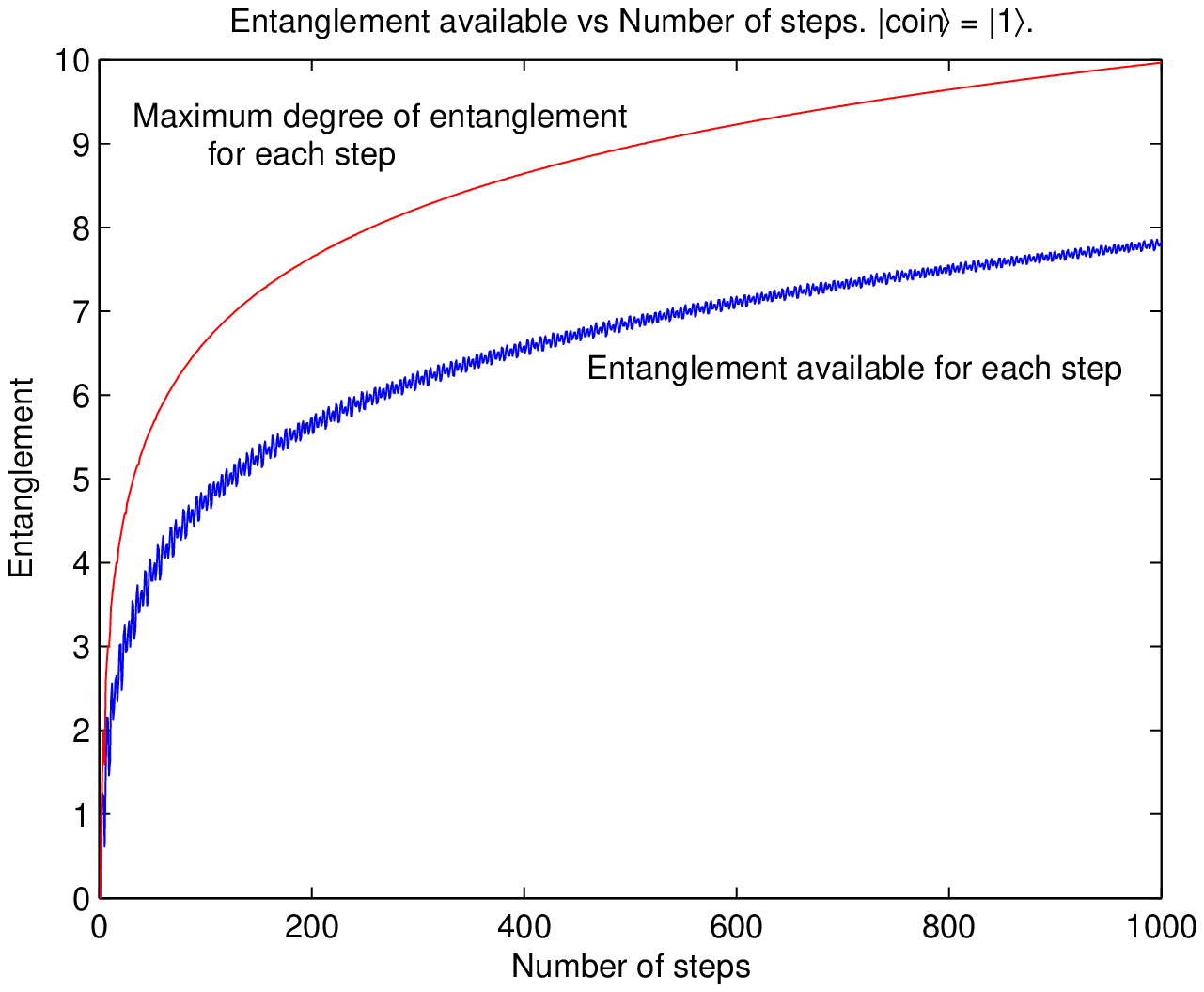}
(ii)\epsfig{width=1.5in,file=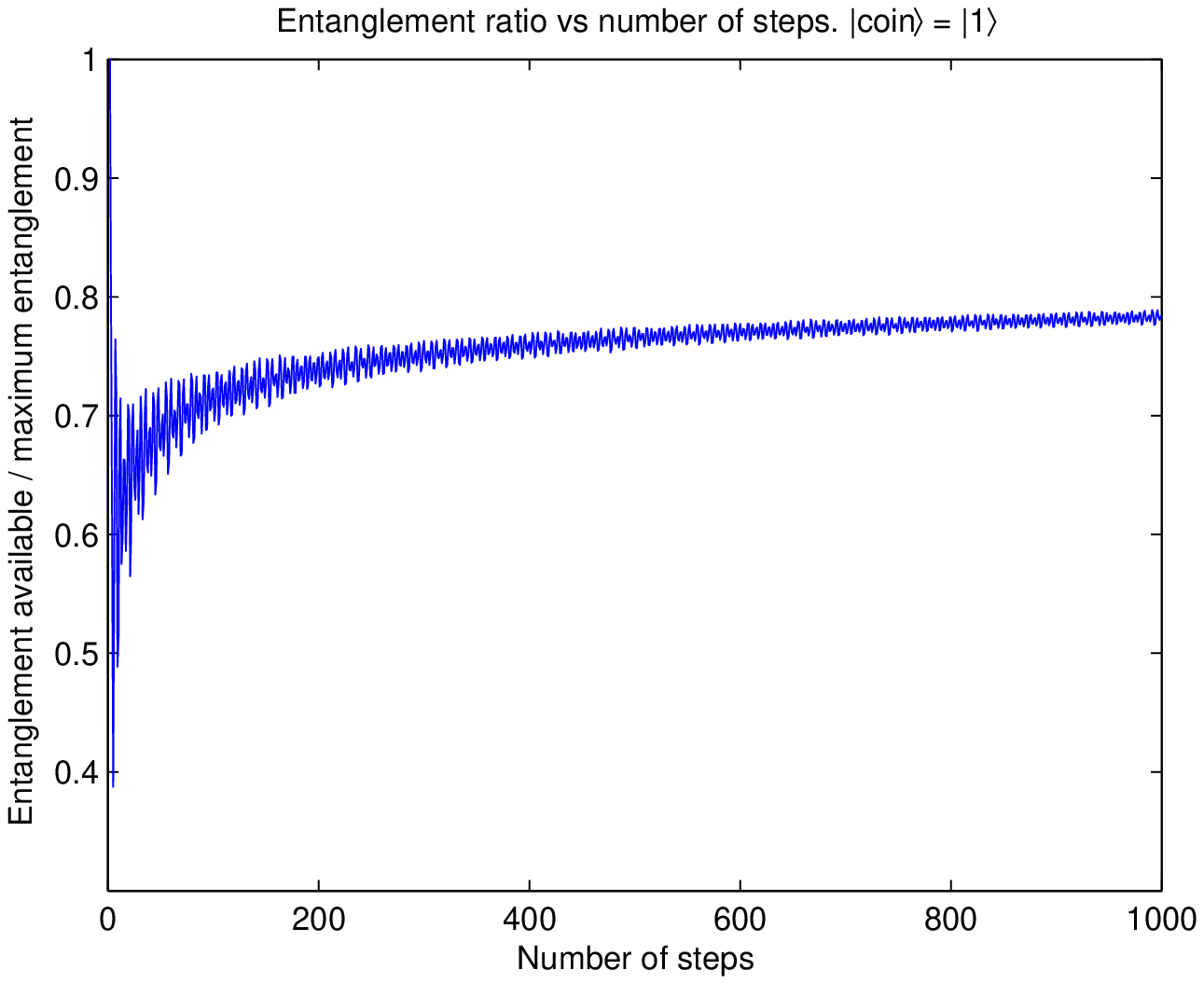}
\caption{{\small Entanglement values for coin post-measurement state
$|\psi\rangle_{t,pm}^{c_1}$ computed from a 1000-steps quantum walk
$|\psi\rangle_{1000} = [{\hat S_{\text{ent}}} (\hat{H} \otimes {\hat I} )]^{1000} |\psi\rangle_0$
with $|\psi\rangle_0$ given by Eq. (\ref{two_initial_condition}),
Eqs. (\ref{hadamard_chapter_qwec_ii}) and (\ref{shift_operator_quant_entanglement})
as coin (${\hat H}$) and shift (${\hat S}$) operators,
and measurement operator ${\hat M}_1$ (Eq. (\ref{observable_qw})).
The thin line of (i) (red color online) shows the maximum degree of entanglement between
walkers attainable in the post-measurement quantum state $|\psi\rangle_{t,pm}^{c_1}$,
and the thick line of (i) (blue color online) shows the actual entanglement between walkers available
at each step. We can see that, asymptotically, the entanglement available
is about 80\% of the corresponding maximum degree of entanglement (plot (ii)).
Note that this amount of entanglement available between walkers (80\%) is {\it less}
than the amount of entanglement available between walkers (90\%) for coin $|1\rangle$
post-measurement quantum state with initial state $|0\rangle_c \otimes |0,0\rangle_p$
(Fig. (\ref{first_condition_coin_one})).
}}\label{second_condition_coin_one}
\end{figure}

\begin{figure}[hbt]
(i)\epsfig{width=1.5in,file=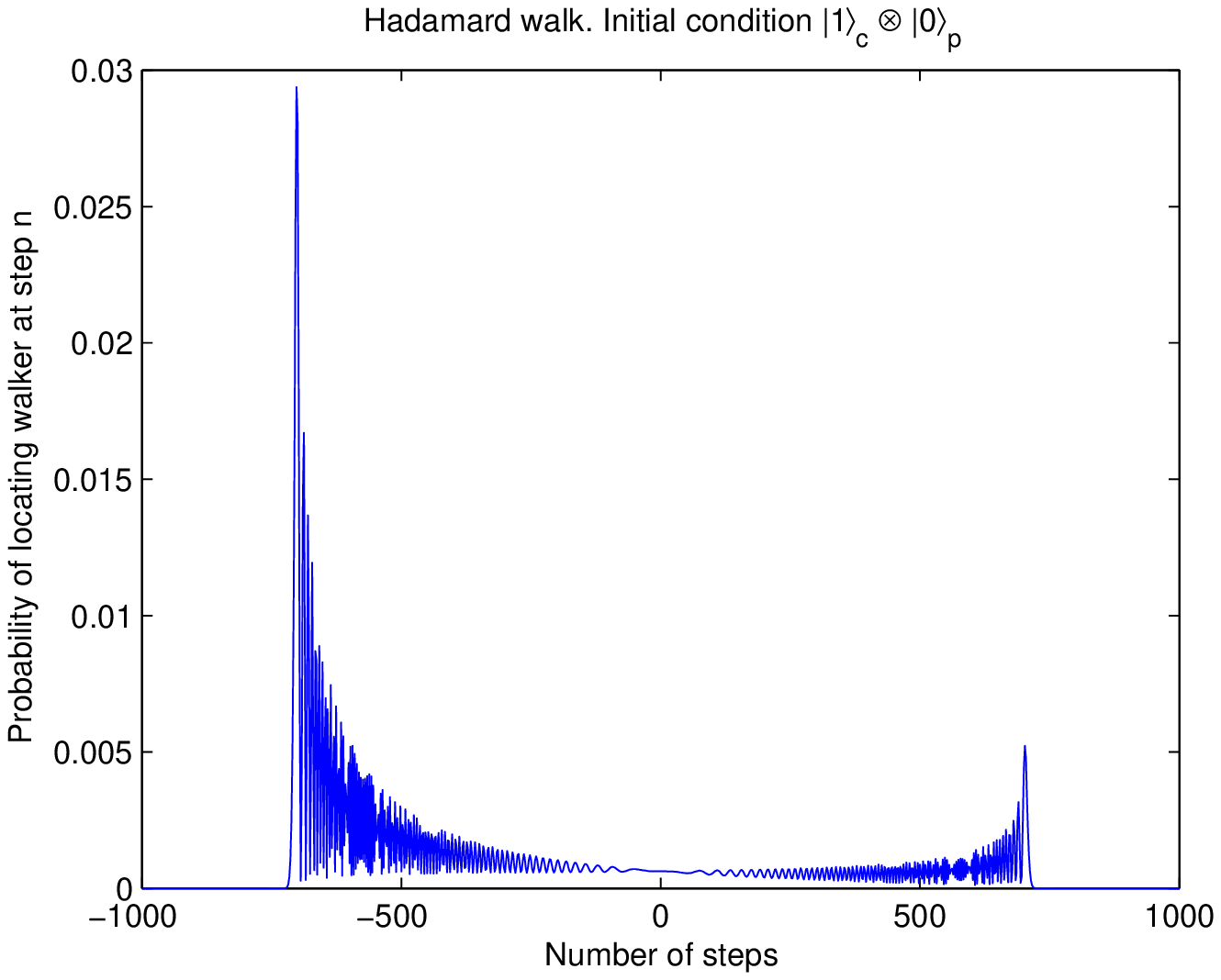}
(ii)\epsfig{width=1.5in,file=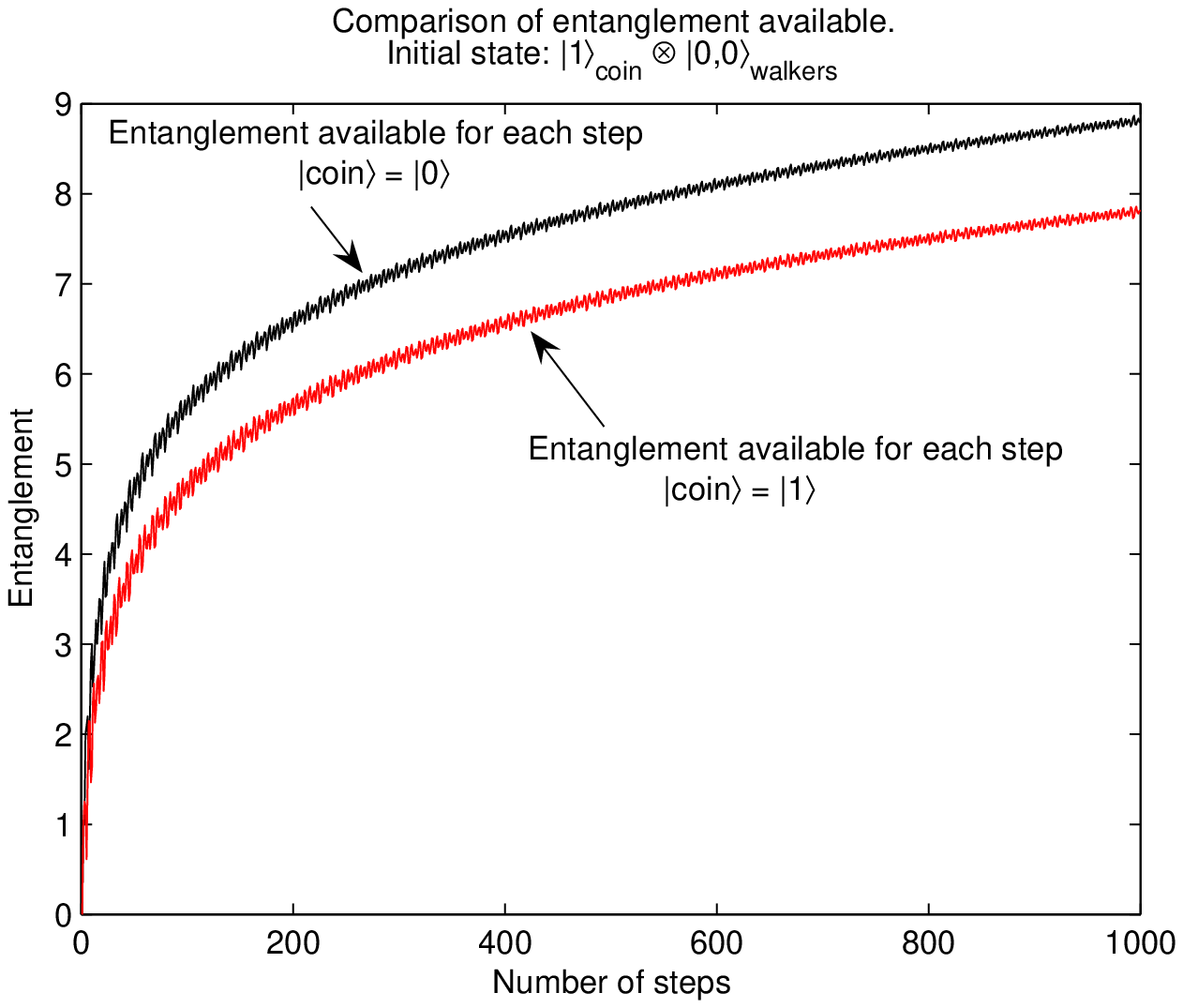}
\caption{{\small Plot (i) presents the probability vs location graph of
a 1000-step Hadamard quantum walk with an initial state
$|1\rangle_c \otimes |0\rangle_p$ and shift operator provided
by Eq. (\ref{shift_single}). The symmetry of this walk, about a line
passing through the origin and perpedicular to the $x$ axis, is the same
as that of a Hadamard quantum walk with initial state given by
$|\psi\rangle = |1\rangle_c \otimes |0,0\rangle_p$ (Eq. (\ref{two_initial_condition}))
and shift operator given by Eq. (\ref{shift_operator_quant_entanglement}).
Plot (ii) is a summary of Figs. (\ref{second_condition_coin_zero}.i)
and (\ref{second_condition_coin_one}.i), and shows that the amount
of entanglement between walkers available in post-measurement state
$|\psi\rangle_{t,pm}^{c_1}$ tends to be {\it less} than the amount of entanglement
between walkers available in post-measurement state $|\psi\rangle_{t,pm}^{c_0}$,
in stark contrast to the numerical results computed for a quantum walk with
total initial state $|0\rangle_c \otimes |0,0\rangle_p$
(Figs. (\ref{first_condition_coin_zero}-\ref{first_condition_entanglement_comparison})).
}}
\label{second_condition_entanglement_comparison}
\end{figure}

\begin{figure}[hbt]
(i)\epsfig{width=1.5in,file=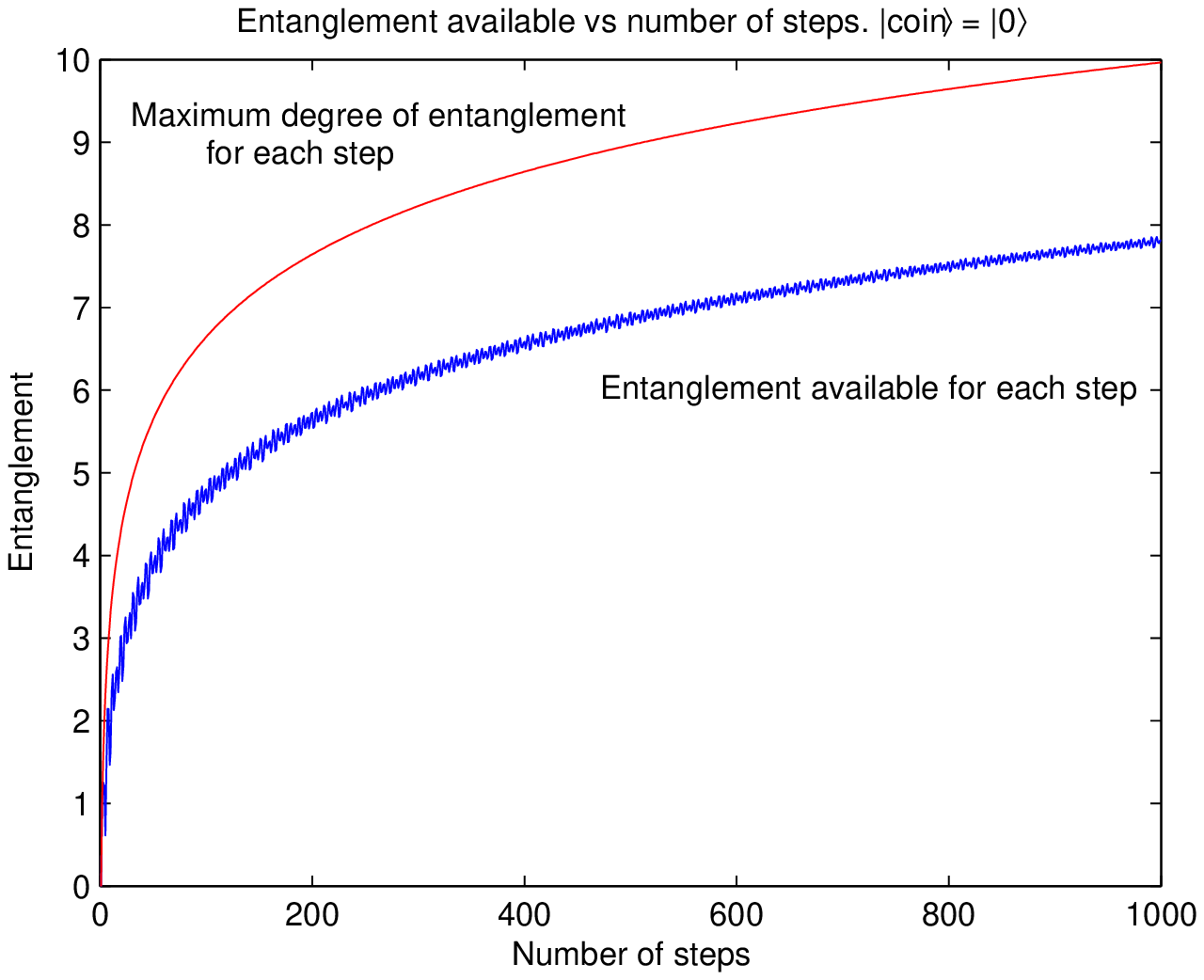}
(ii)\epsfig{width=1.5in,file=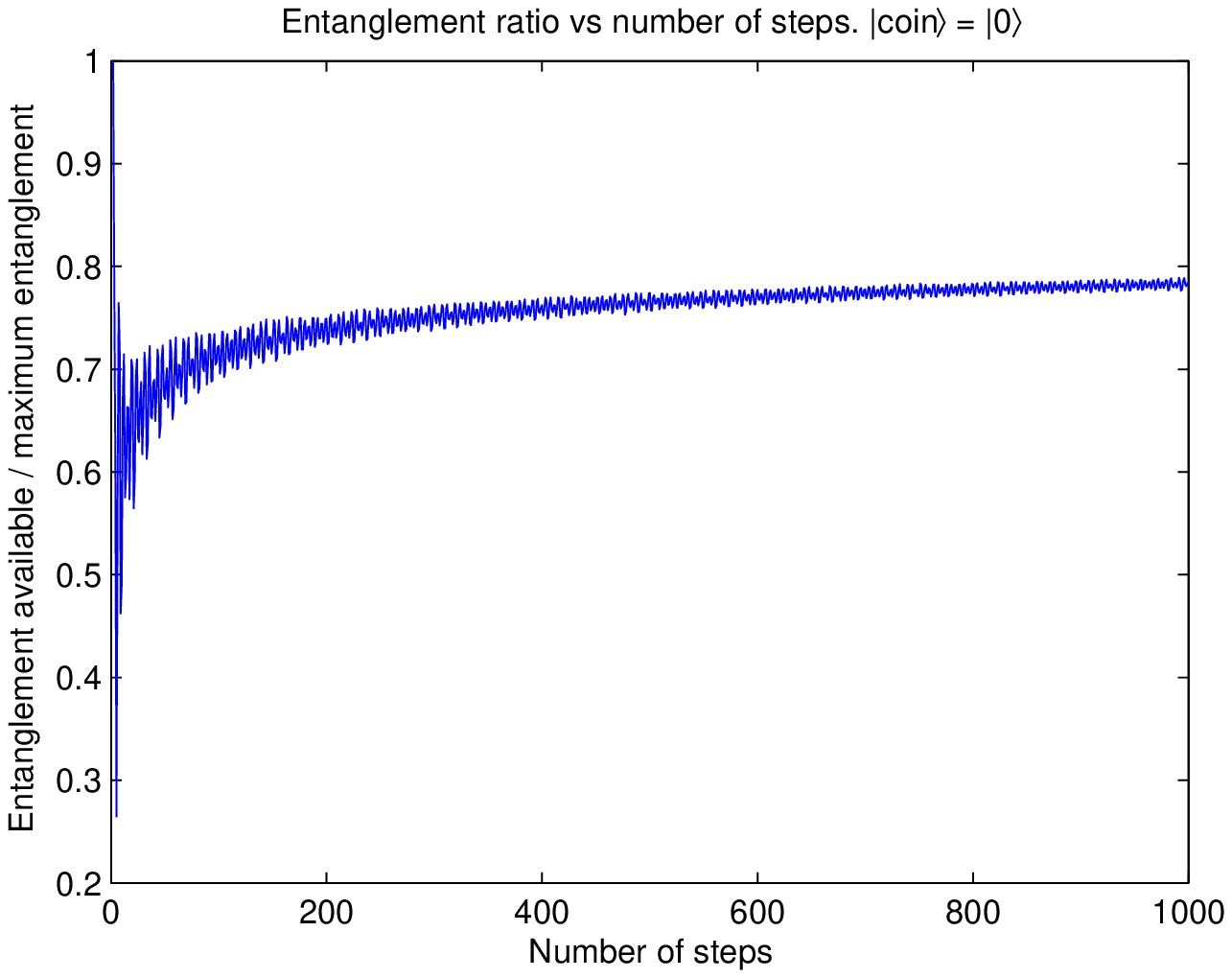}
\caption{{\small Entanglement values for coin $|0\rangle_c$ post-measurement state
$|\psi\rangle_{t,pm}^{c_0}$ computed from a 1000-steps quantum walk
$|\psi\rangle_{1000} = [{\hat S_{\text{ent}}} (\hat{H} \otimes {\hat I} )]^{1000} |\psi\rangle_0$
with $|\psi\rangle_0 = ({1 \over \sqrt{2}}|0\rangle_c + {i \over \sqrt{2}}|1\rangle_c) \otimes |0,0\rangle_p$
given by Eq. (\ref{three_initial_condition}), coin (${\hat H}$) and shift (${\hat S}$) operators given by
Eqs. (\ref{hadamard_chapter_qwec_ii}) and (\ref{shift_operator_quant_entanglement})
respectively, and measurement operator ${\hat M}_0$ (Eq. (\ref{observable_qw})).
The thin line of (i) (red color online) shows the maximum degree of entanglement between
walkers attainable in the post-measurement quantum state $|\psi\rangle_{t,pm}^{c_0}$,
and the thick line of (i) (blue color online) shows the actual entanglement between walkers available
at each step. The asymptotical behavior of entanglement values for this quantum walk
is the same as that shown by a quantum walk with total initial state
$|0\rangle_c \otimes |0,0\rangle_p$ (Fig. (\ref{first_condition_coin_zero})).
}}
\label{third_condition_coin_zero}
\end{figure}

\begin{figure}[hbt]
(i)\epsfig{width=1.5in,file=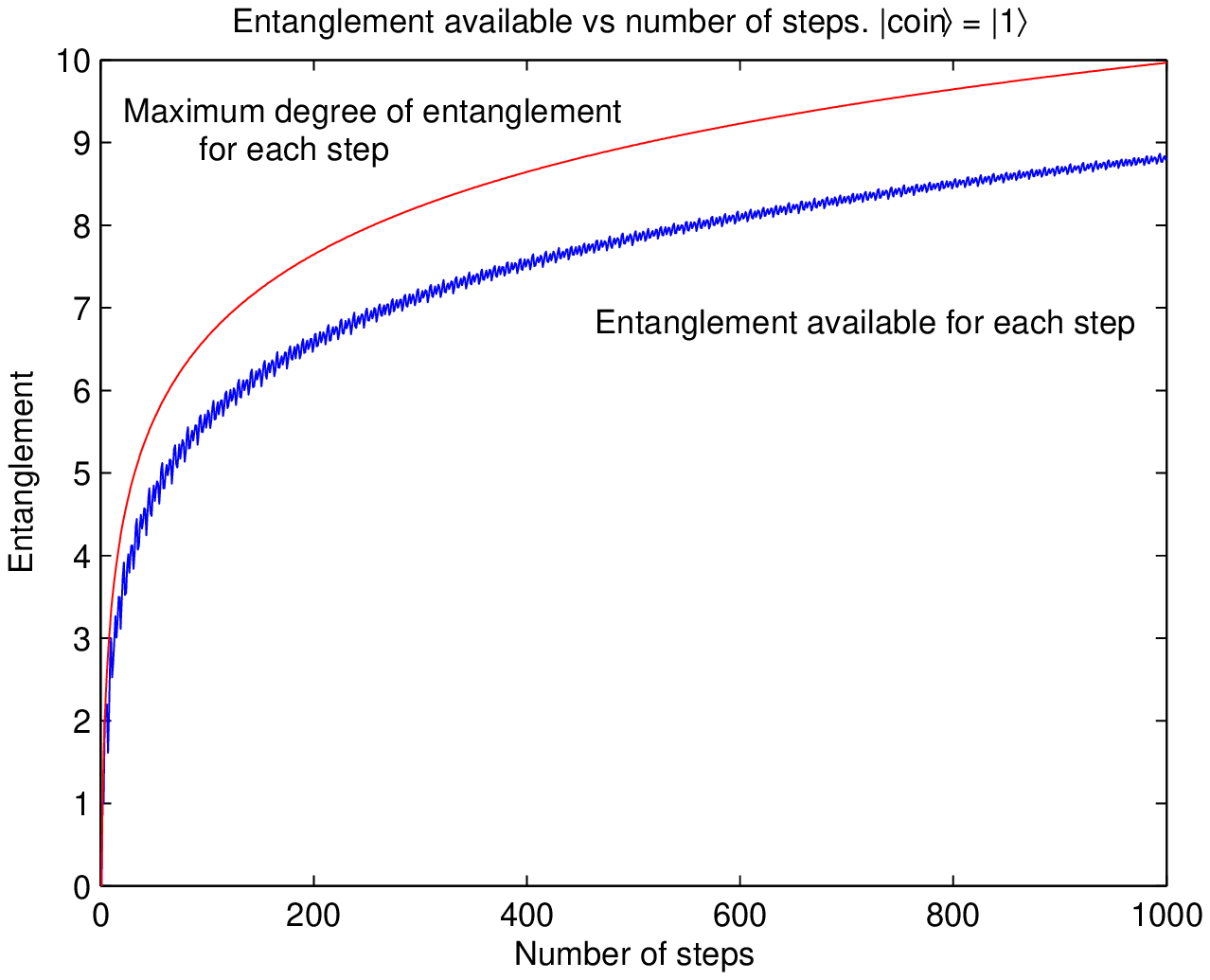}
(ii)\epsfig{width=1.5in,file=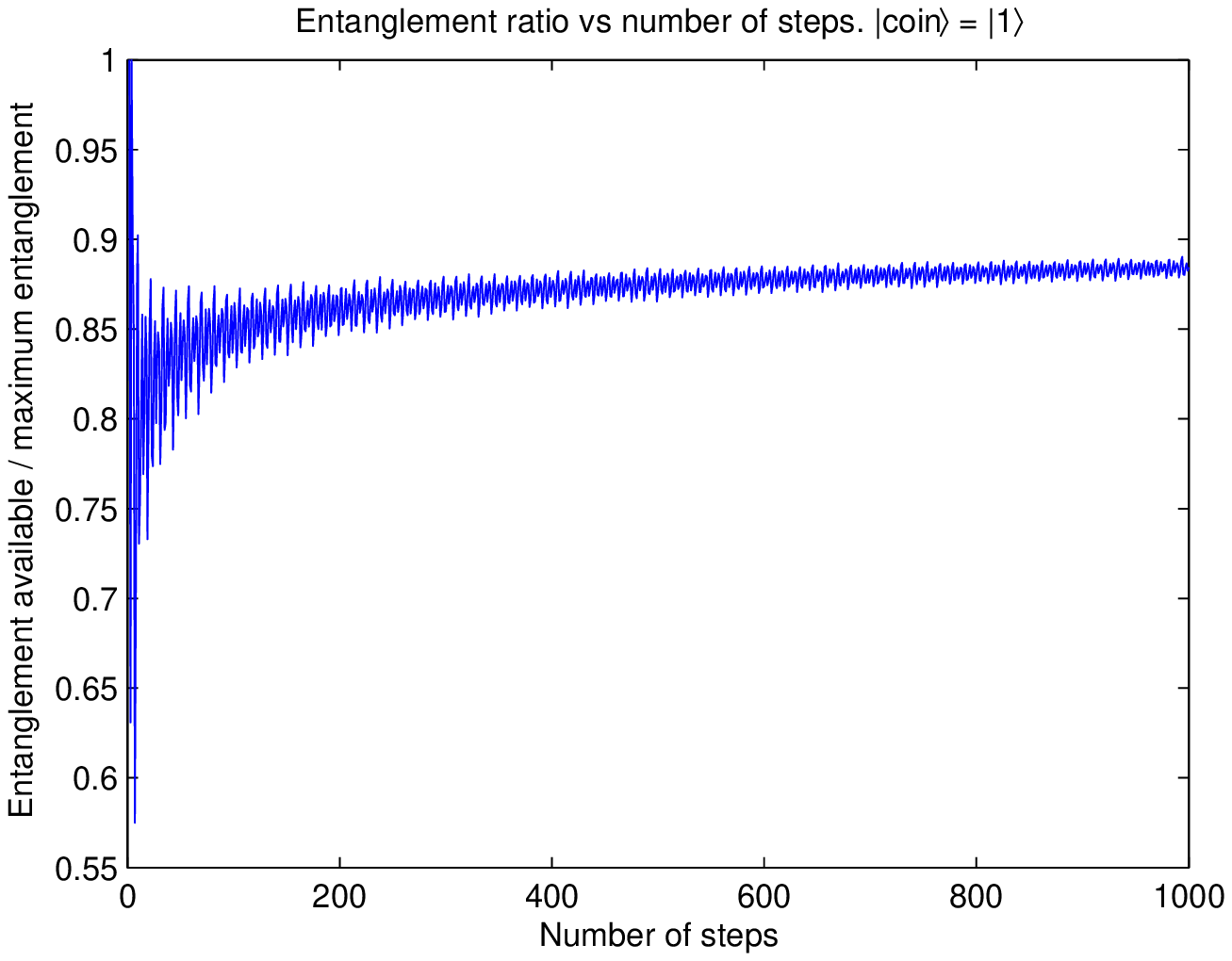}
\caption{{\small Entanglement values for coin $|1\rangle_c$ post-measurement state
$|\psi\rangle_{t,pm}^{c_1}$ computed from a 1000-steps quantum walk
$|\psi\rangle_{1000} = [{\hat S_{\text{ent}}} (\hat{H} \otimes {\hat I} )]^{1000} |\psi\rangle_0$
with $|\psi\rangle_0 = ({1 \over \sqrt{2}}|0\rangle_c + {i \over \sqrt{2}}|1\rangle_c) \otimes |0,0\rangle_p$
given by Eq. (\ref{three_initial_condition}), coin (${\hat H}$) and shift (${\hat S}$) operators given by
Eqs. (\ref{hadamard_chapter_qwec_ii}) and (\ref{shift_operator_quant_entanglement})
respectively, and measurement operator ${\hat M}_1$ (Eq. (\ref{observable_qw})).
The thin line of (i) (red color online) shows the maximum degree of entanglement between
walkers attainable in the post-measurement quantum state $|\psi\rangle_{t,pm}^{c_1}$,
and the thick line of (i) (blue color online) shows the actual entanglement between walkers available
at each step. The asymptotical behavior of entanglement values for this quantum walk
is the same as that shown by a quantum walk with total initial state
$|0\rangle_c \otimes |0,0\rangle_p$ (Fig. (\ref{first_condition_coin_one})).
}}
\label{third_condition_coin_one}
\end{figure}

\begin{figure}[hbt]
(i)\epsfig{width=1.5in,file=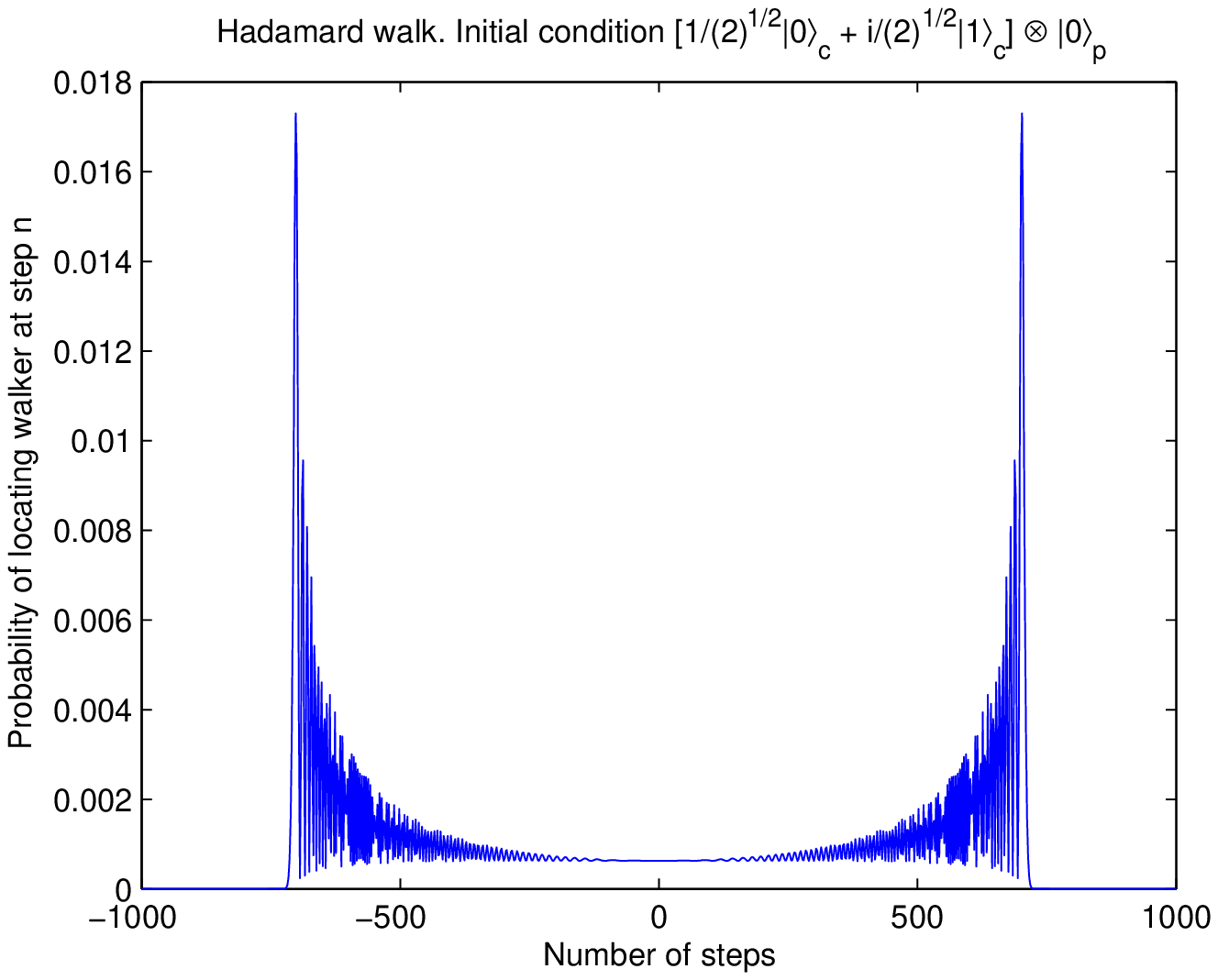}
(ii)\epsfig{width=1.5in,file=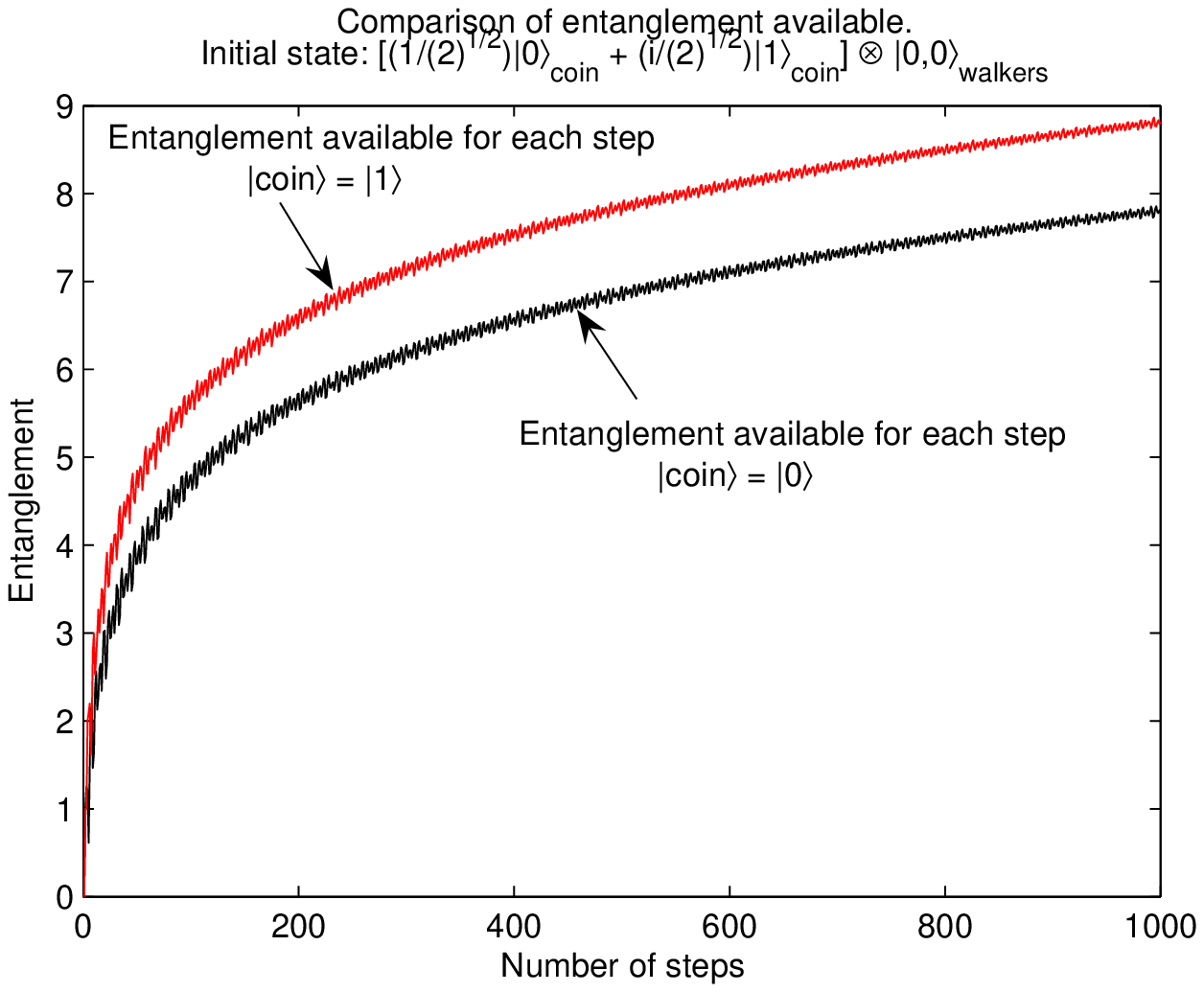}
\caption{{\small Plot (i) presents the probability vs location graph
of a 1000-step Hadamard quantum walk with an initial state
$({1 \over \sqrt{2}}|0\rangle_c + {i \over \sqrt{2}}|1\rangle_c) \otimes |0\rangle_p$
and shift operator provided by Eq. (\ref{shift_single}). The symmetry
of the probability distribution shown in plot (i) is the same
as that of a Hadamard quantum walk with initial state given by
$|\psi\rangle_0 = ({1 \over \sqrt{2}}|0\rangle_c + {i \over \sqrt{2}}|1\rangle_c) \otimes |0,0\rangle_p$ and
shift operator given by Eq. (\ref{shift_operator_quant_entanglement}).
Although the symmetry of plot (i) is significantly different from
that of Fig. (\ref{first_condition_entanglement_comparison}.i),
plot (ii) shows the same asymptotical behavior as that of
Fig. (\ref{first_condition_entanglement_comparison}.ii).
}}
\label{third_condition_entanglement_comparison}
\end{figure}

We now focus on Figs. (\ref{second_condition_coin_zero}), (\ref{second_condition_coin_one})
and (\ref{second_condition_entanglement_comparison}), which present the numerical behavior
of a quantum walk with initial quantum state given by Eq. (\ref{two_initial_condition}),
and Eqs. (\ref{hadamard_chapter_qwec_ii}) and (\ref{shift_operator_quant_entanglement})
as coin and shift operators, respectively.

As in the previous case, Figs. (\ref{second_condition_coin_zero})
and (\ref{second_condition_coin_one}) display the results of measuring
entanglement between walkers in a coin $|0\rangle_c$ post-measurement
state $|\psi\rangle_{t,pm}^{c_0}$ and a coin $|1\rangle_c$ post-measurement
state $|\psi\rangle_{t,pm}^{c_1}$. However, and in contrast to
Figs. (\ref{first_condition_coin_zero})-(\ref{first_condition_entanglement_comparison}),
in this case we see that, as the number of steps increases, {\it the entanglement between walkers
for $|\psi\rangle_{t,pm}^{c_0}$ (about 90\% with respect to the degree
of entanglement attainable in each step, Fig. (\ref{second_condition_coin_zero}.ii))
is} {\bf higher} {\it than that of state $|\psi\rangle_{t,pm}^{c_1}$ (about 80\% with respect
to the degree of entanglement attainable in each step, Fig. (\ref{second_condition_coin_one}.ii)).}
As we can see by comparing Figs. (\ref{first_condition_entanglement_comparison}) and
(\ref{second_condition_entanglement_comparison}),
the symmetry of the probability distribution computed with initial
quantum state given by Eq. (\ref{two_initial_condition}) (Fig. (\ref{second_condition_entanglement_comparison}.i))
seems to have a significant effect on the actual entanglement values for
$|\psi\rangle_{t,pm}^{c_0}$ and $|\psi\rangle_{t,pm}^{c_1}$.

So, a natural step forward is to compute quantum walks with initial states
that produce symmetric probability distributions, in order to see the
asymptotical behavior of entanglement. With this thought in mind
we have computed the following three sets of numerical simulations.

The first set consists of Figs. (\ref{third_condition_coin_zero}), (\ref{third_condition_coin_one})
and (\ref{third_condition_entanglement_comparison}), in which we expose the numerical behavior
of a quantum walk with initial quantum state given by Eq. (\ref{three_initial_condition}),
i.e. $|\psi\rangle_0 = ({1 \over \sqrt{2}}|0\rangle_c + {i \over \sqrt{2}}|1\rangle_c) \otimes |0,0\rangle_p$,
and Eqs. (\ref{hadamard_chapter_qwec_ii}) and (\ref{shift_operator_quant_entanglement})
as coin and shift operators, respectively.
Fig. (\ref{third_condition_coin_zero}) shows the results of measuring entanglement
between walkers in a coin $|0\rangle_c$ post-measurement state $|\psi\rangle_{t,pm}^{c_0}$,
while Fig. (\ref{third_condition_coin_one}) introduces corresponding results for
a coin $|1\rangle_c$ post-measurement state $|\psi\rangle_{t,pm}^{c_1}$.

\begin{figure}[hbt]
(i)\epsfig{width=1.5in,file=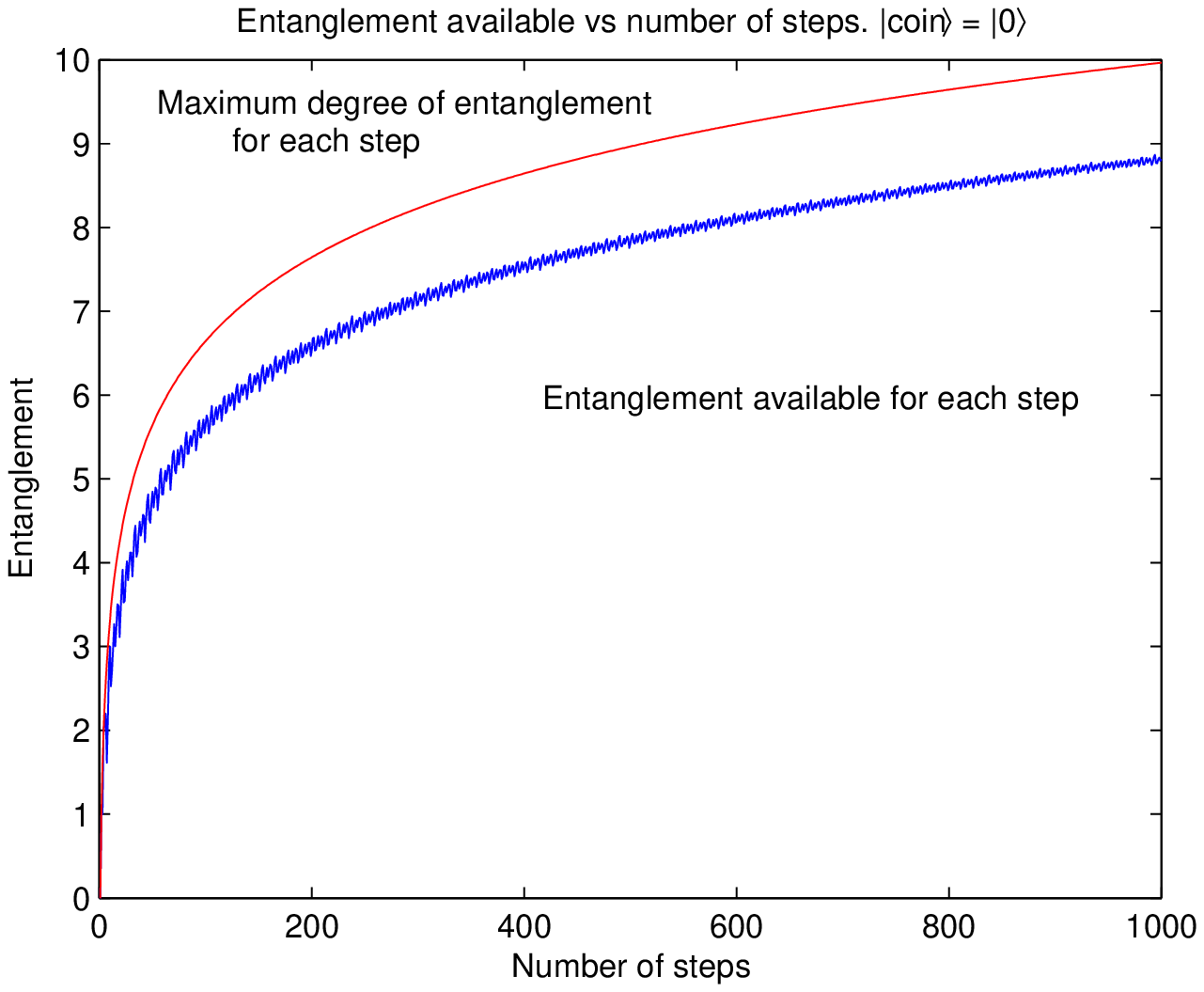}
(ii)\epsfig{width=1.5in,file=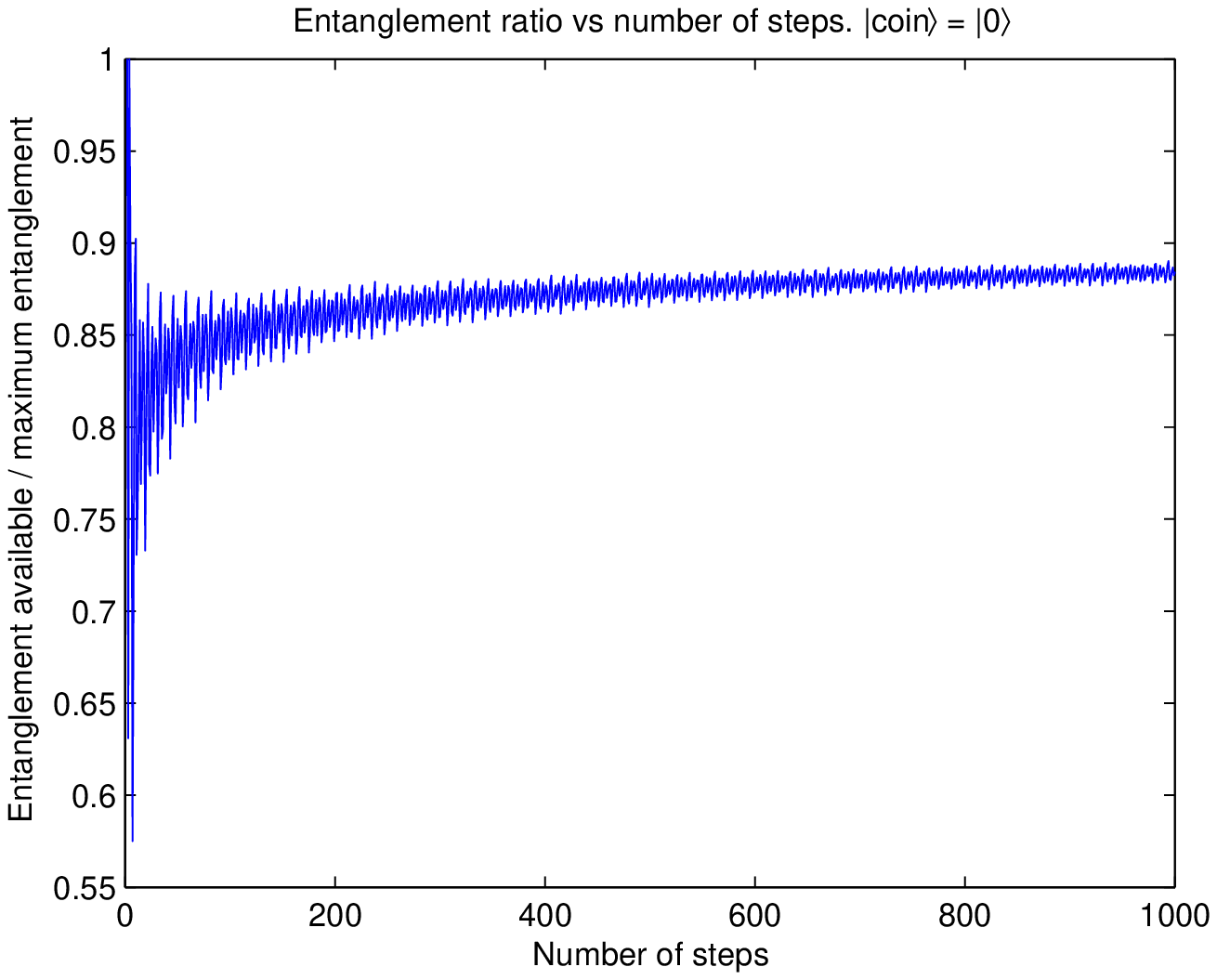}
\caption{{\small Entanglement values for coin $|0\rangle_c$ post-measurement state
$|\psi\rangle_{t,pm}^{c_0}$ computed from a 1000-steps quantum walk
$|\psi\rangle_{1000} = [{\hat S_{\text{ent}}} (\hat{H} \otimes {\hat I} )]^{1000} |\psi\rangle_0$
with $|\psi\rangle_0 = ({i \over \sqrt{2}}|0\rangle_c + {1 \over \sqrt{2}}|1\rangle_c) \otimes |0,0\rangle_p$
given by Eq. (\ref{fourth_initial_condition}), coin (${\hat H}$) and shift (${\hat S}$) operators given by
Eqs. (\ref{hadamard_chapter_qwec_ii}) and (\ref{shift_operator_quant_entanglement})
respectively, and measurement operator ${\hat M}_0$ (Eq. (\ref{observable_qw})).
The thin line of (i) (red color online) shows the maximum degree of entanglement between
walkers attainable in the post-measurement quantum state $|\psi\rangle_{t,pm}^{c_0}$,
and the thick line of (i) (blue color online) shows the actual entanglement between walkers available
at each step. The asymptotical behavior of entanglement values for this quantum walk
is the same as that shown by a quantum walk with total initial state
$|1\rangle_c \otimes |0,0\rangle_p$ (Fig. (\ref{second_condition_coin_zero})).
}}
\label{fourth_condition_coin_zero}
\end{figure}

\begin{figure}[hbt]
(i)\epsfig{width=1.5in,file=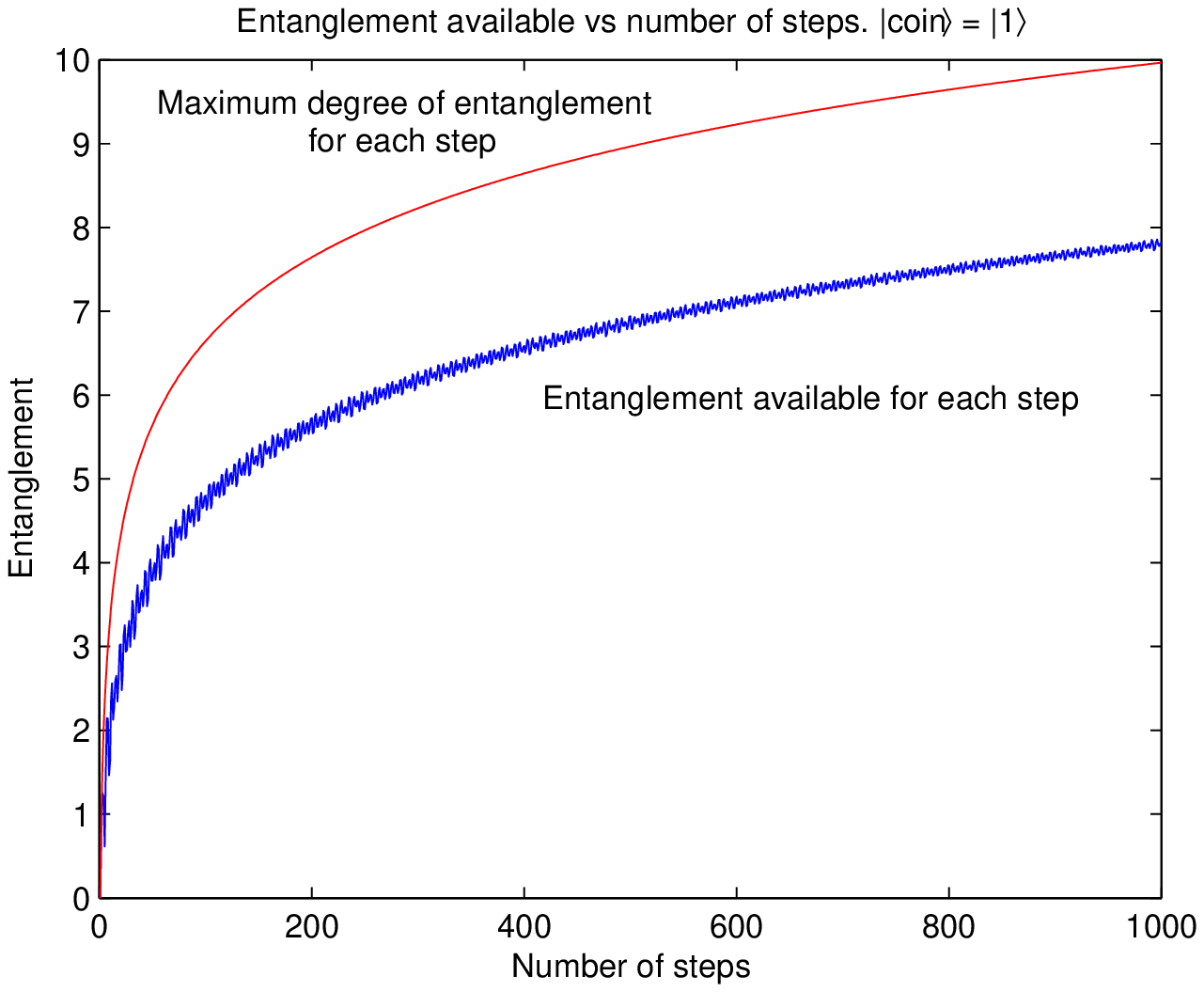}
(ii)\epsfig{width=1.5in,file=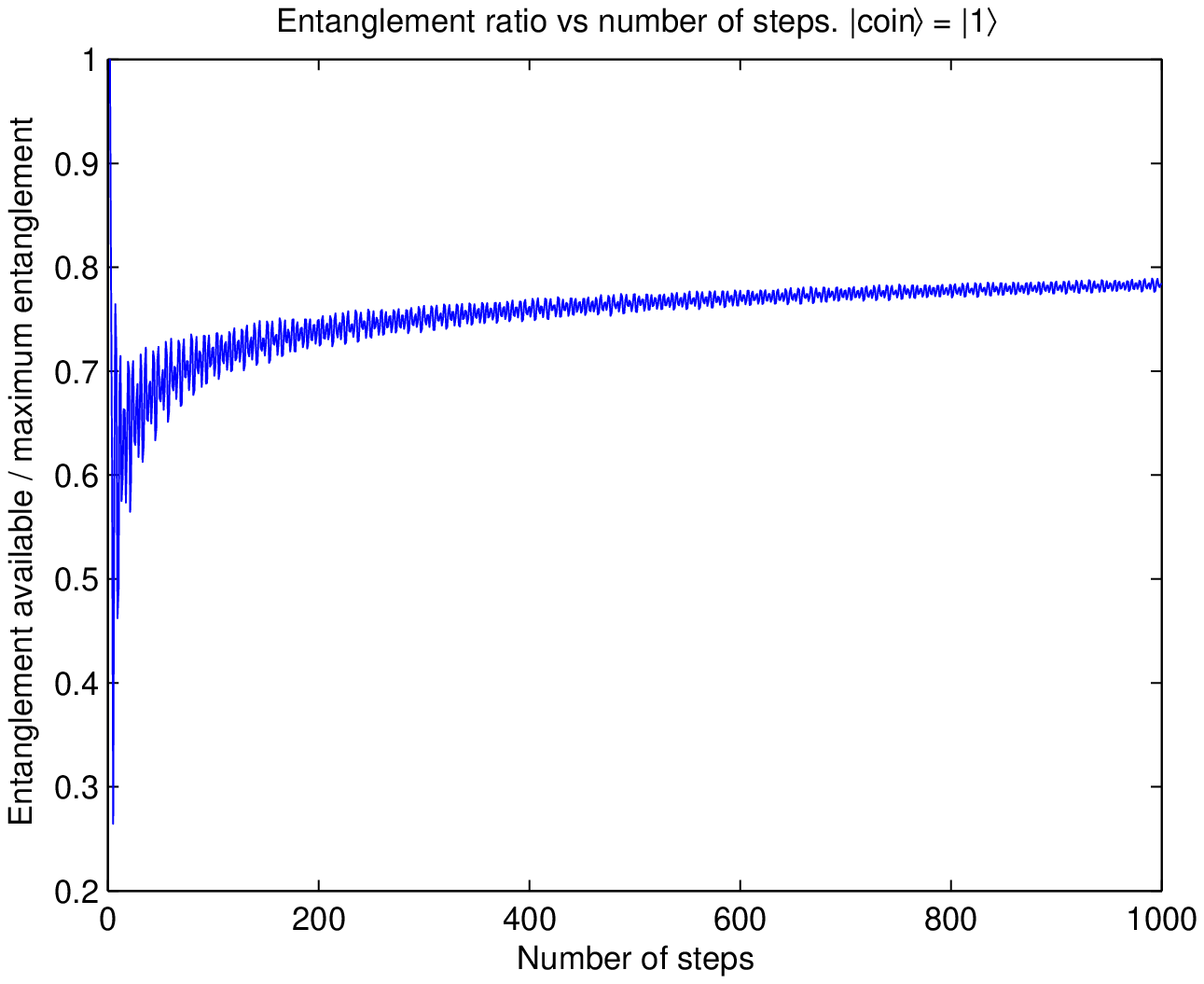}
\caption{{\small Entanglement values for coin $|1\rangle_c$ post-measurement state
$|\psi\rangle_{t,pm}^{c_1}$ computed from a 1000-steps quantum walk
$|\psi\rangle_{1000} = [{\hat S_{\text{ent}}} (\hat{H} \otimes {\hat I} )]^{1000} |\psi\rangle_0$
with $|\psi\rangle_0 = ({i \over \sqrt{2}}|0\rangle_c + {1 \over \sqrt{2}}|1\rangle_c) \otimes |0,0\rangle_p$
given by Eq. (\ref{fourth_initial_condition}),  coin (${\hat H}$) and shift (${\hat S}$) operators given by
Eqs. (\ref{hadamard_chapter_qwec_ii}) and (\ref{shift_operator_quant_entanglement})
respectively, and measurement operator ${\hat M}_1$ (Eq. (\ref{observable_qw})).
The thin line of (i) (red color online) shows the maximum degree of entanglement between
walkers attainable in the post-measurement quantum state $|\psi\rangle_{t,pm}^{c_1}$,
and the thick line of (i) (blue color online) shows the actual entanglement between walkers available
at each step. The asymptotical behavior of entanglement values for this quantum walk
is the same as that shown by a quantum walk with total initial state
$|1\rangle_c \otimes |0,0\rangle_p$ (Fig. (\ref{second_condition_coin_one})).
}}
\label{fourth_condition_coin_one}
\end{figure}

\begin{figure}[hbt]
(i)\epsfig{width=1.5in,file=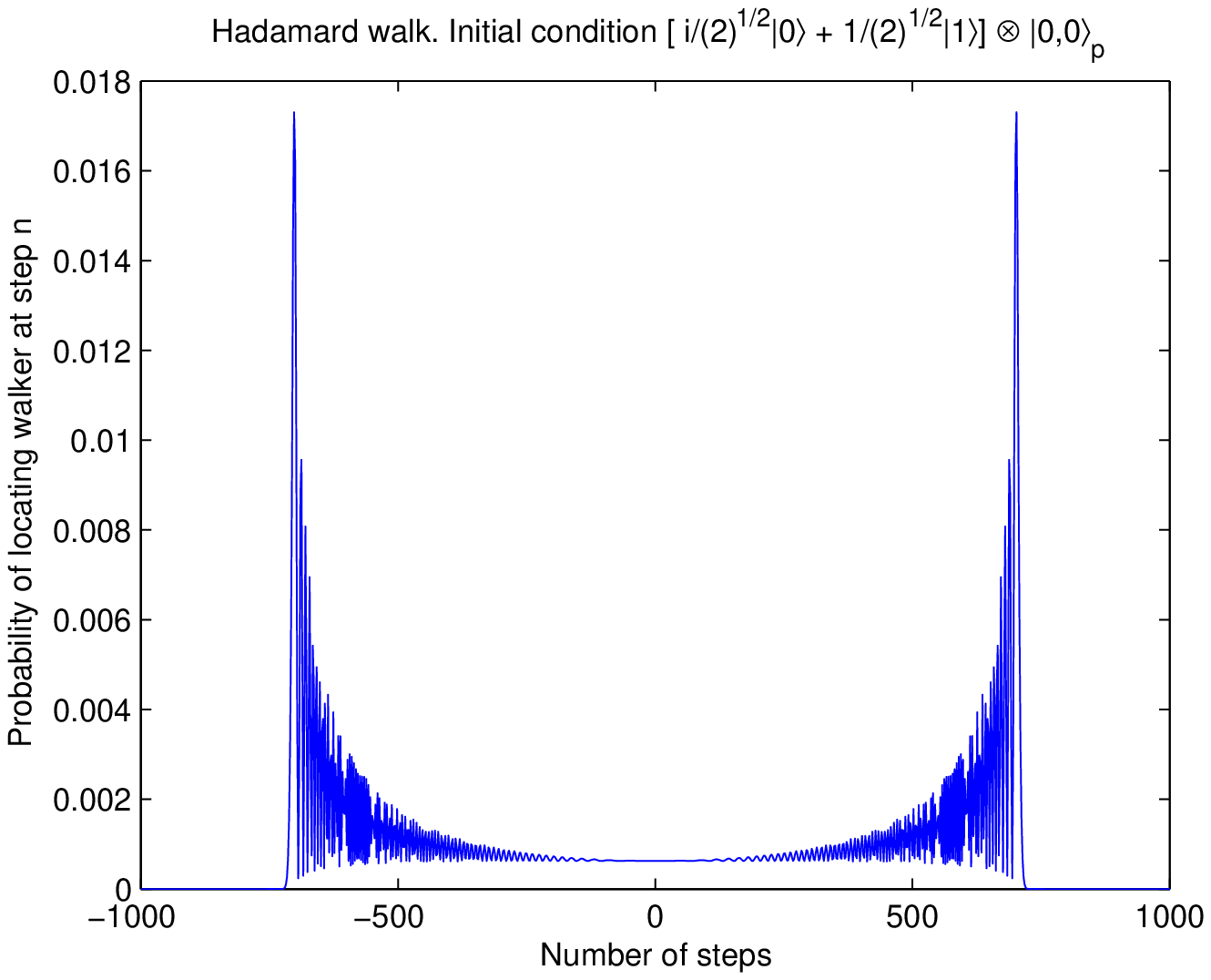}
(ii)\epsfig{width=1.5in,file=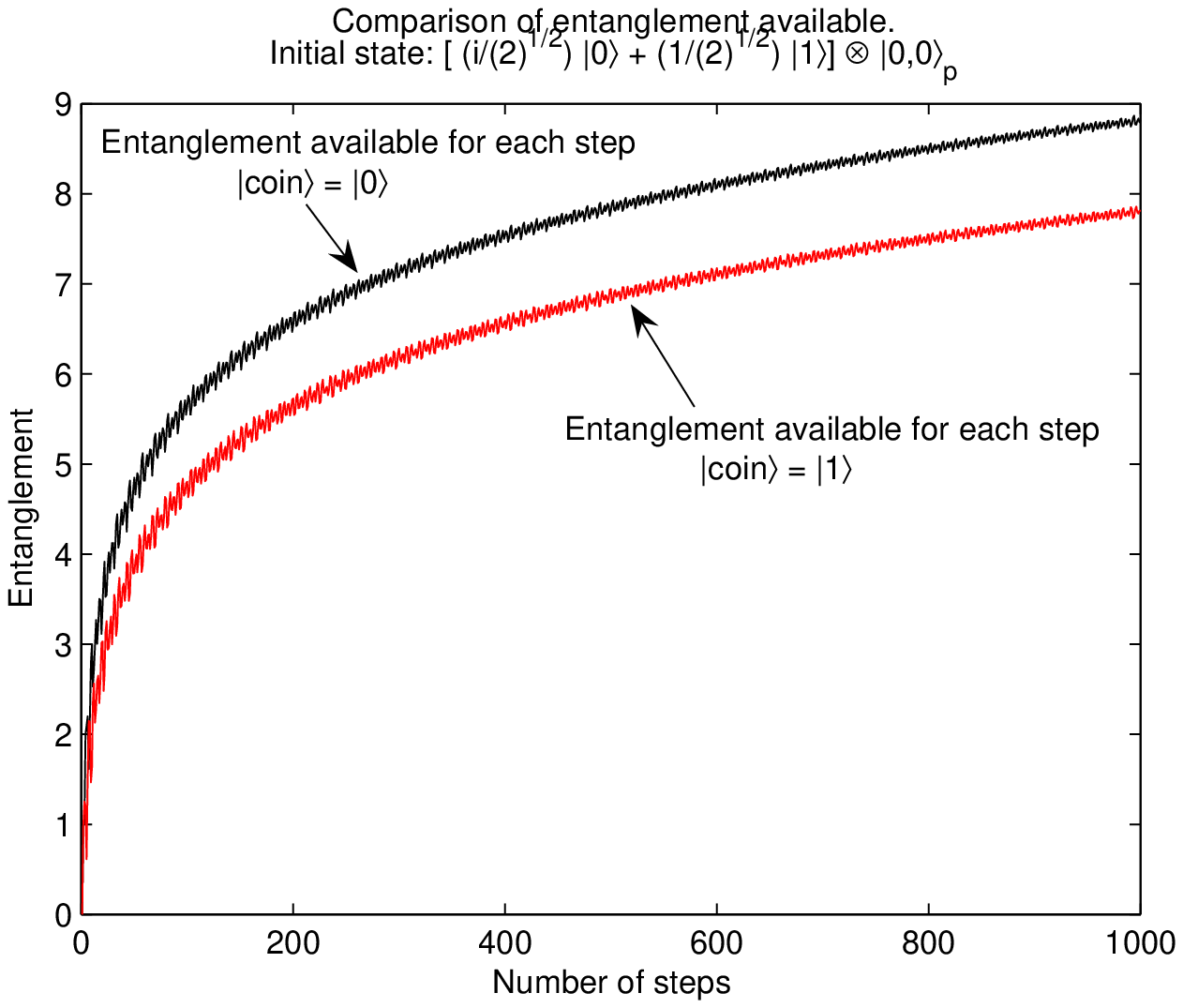}
\caption{{\small Plot (i) presents the probability vs location graph
of a 1000-step Hadamard quantum walk with an initial state
$({i \over \sqrt{2}}|0\rangle_c + {1 \over \sqrt{2}}|1\rangle_c) \otimes |0\rangle_p$
and shift operator provided by Eq. (\ref{shift_single}). The symmetry
of the probability distribution shown in plot (i) is the same
as that of a Hadamard quantum walk with initial state given by
$|\psi\rangle = ({i \over \sqrt{2}}|0\rangle_c + {1 \over \sqrt{2}}|1\rangle_c) \otimes |0,0\rangle_p$ and
shift operator given by Eq. (\ref{shift_operator_quant_entanglement}).
Although the symmetry of plot (i) is significantly different from
that of Fig. (\ref{second_condition_entanglement_comparison}.i),
plot (ii) shows the same asymptotical behavior as that of
Fig. (\ref{second_condition_entanglement_comparison}.ii).
}}
\label{fourth_condition_entanglement_comparison}
\end{figure}

Although an initial quantum state of the form given by Eq. (\ref{three_initial_condition})
produces a balanced probability distribution (Fig. (\ref{third_condition_entanglement_comparison}.i)),
such a property does not have a significant effect on the degree of
entanglement between walkers (Fig. (\ref{third_condition_entanglement_comparison}.ii)).
In fact, comparing plots from Figs. (\ref{first_condition_entanglement_comparison}.ii)
and (\ref{third_condition_entanglement_comparison}.ii) shows that the
asymptotical behavior of entanglement values for a quantum walk with initial
state  given by Eq. (\ref{zero_initial_condition}) is the same as those entanglement values
computed for a quantum walk with initial state given by Eq. (\ref{three_initial_condition}).

Figs. (\ref{fourth_condition_coin_zero}) - (\ref{fourth_condition_entanglement_comparison})
introduce the asymptotics of entanglement values for a quantum walk with initial
state given by Eq. (\ref{fourth_initial_condition}).
Again, although the initial state $|\psi\rangle_0 = ({i \over \sqrt{2}}|0\rangle_c + {1 \over \sqrt{2}}|1\rangle_c) \otimes |0,0\rangle_p$
produces a symmetrical probability distribution (Fig. (\ref{fourth_condition_entanglement_comparison}.i)),
we notice that the asymptotical behavior of entanglement values for a
coin $|0\rangle$ post-measurement quantum state $|\psi\rangle_{t,pm}^{c_0}$
is different from that of a
coin $|1\rangle$ post-measurement quantum state $|\psi\rangle_{t,pm}^{c_1}$
(Fig. (\ref{fourth_condition_entanglement_comparison}.ii)).
In fact, comparing plots from Figs. (\ref{second_condition_coin_zero}) and
(\ref{fourth_condition_coin_zero}) for a coin
$|0\rangle$ post-measurement quantum state $|\psi\rangle_{t,pm}^{c_0}$,
and plots from Figs. (\ref{second_condition_coin_one}) and
(\ref{fourth_condition_coin_one}) for a coin
$|1\rangle$ post-measurement quantum state $|\psi\rangle_{t,pm}^{c_1}$,
shows that the asymptotics of entanglement values for initial states given
by Eqs. (\ref{two_initial_condition}) and (\ref{fourth_initial_condition})
are the same.

\begin{figure}[hbt]
(i)\epsfig{width=1.5in,file=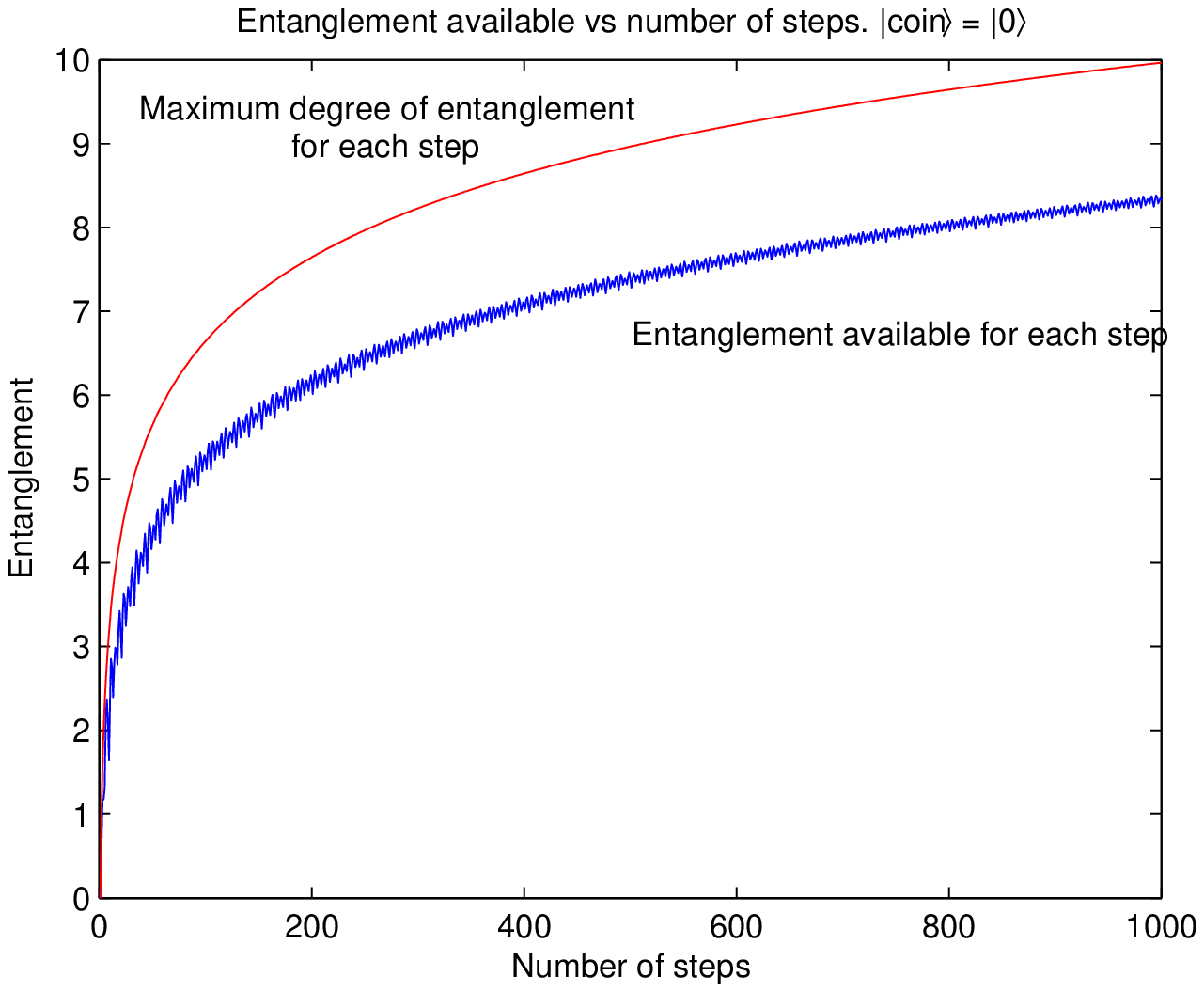}
(ii)\epsfig{width=1.5in,file=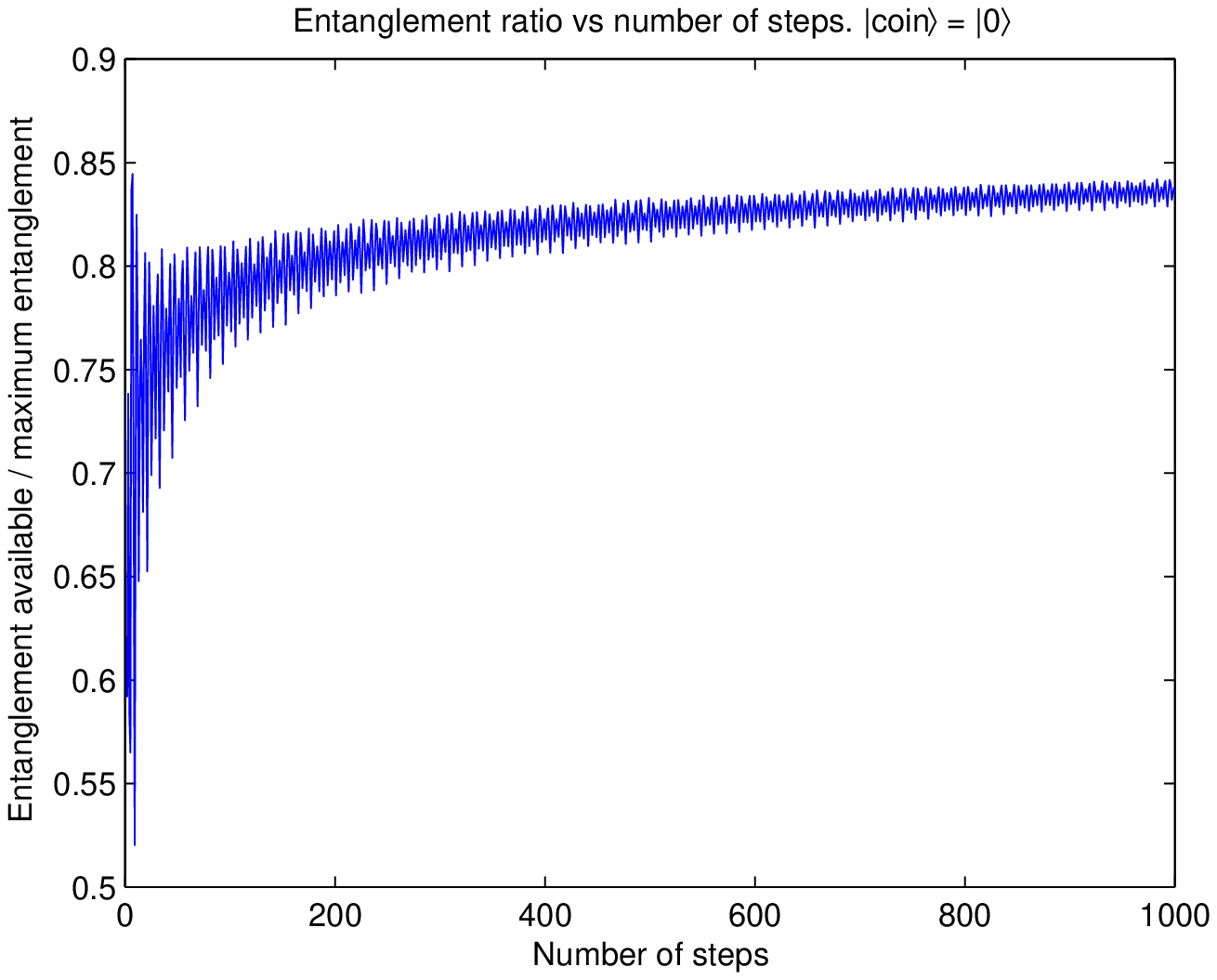}
\caption{{\small Entanglement values for coin $|0\rangle_c$ post-measurement state
$|\psi\rangle_{t,pm}^{c_0}$ computed from a 1000-steps quantum walk
$|\psi\rangle_{1000} = [{\hat S_{\text{ent}}} (\hat{H} \otimes {\hat I} )]^{1000} |\psi\rangle_0$
with $|\psi\rangle_0 = (\sqrt{0.85}|0\rangle_c - \sqrt{0.15}|1\rangle_c) \otimes |0,0\rangle_p$
given by Eq. (\ref{five_initial_condition}), coin (${\hat H}$) and shift (${\hat S}$) operators given by
Eqs. (\ref{hadamard_chapter_qwec_ii}) and (\ref{shift_operator_quant_entanglement})
respectively, and measurement operator ${\hat M}_0$ (Eq. (\ref{observable_qw})).
The thin line of (i) (red color online) shows the maximum degree of entanglement between
walkers attainable in the post-measurement quantum state $|\psi\rangle_{t,pm}^{c_0}$,
and the thick line of (i) (blue color online) shows the actual entanglement between walkers available
at each step. We can see that the asymptotics of entanglement values given in
plot (ii) tend to the same values as those shown in Fig. (\ref{fifth_condition_coin_one}),
obtained from a coin $|1\rangle_c$ post-measurement state $|\psi\rangle_{t,pm}^{c_1}$
computed from a quantum walk with the same initial state (Eq. (\ref{five_initial_condition})).
}}
\label{fifth_condition_coin_zero}
\end{figure}

\begin{figure}[hbt]
(i)\epsfig{width=1.5in,file=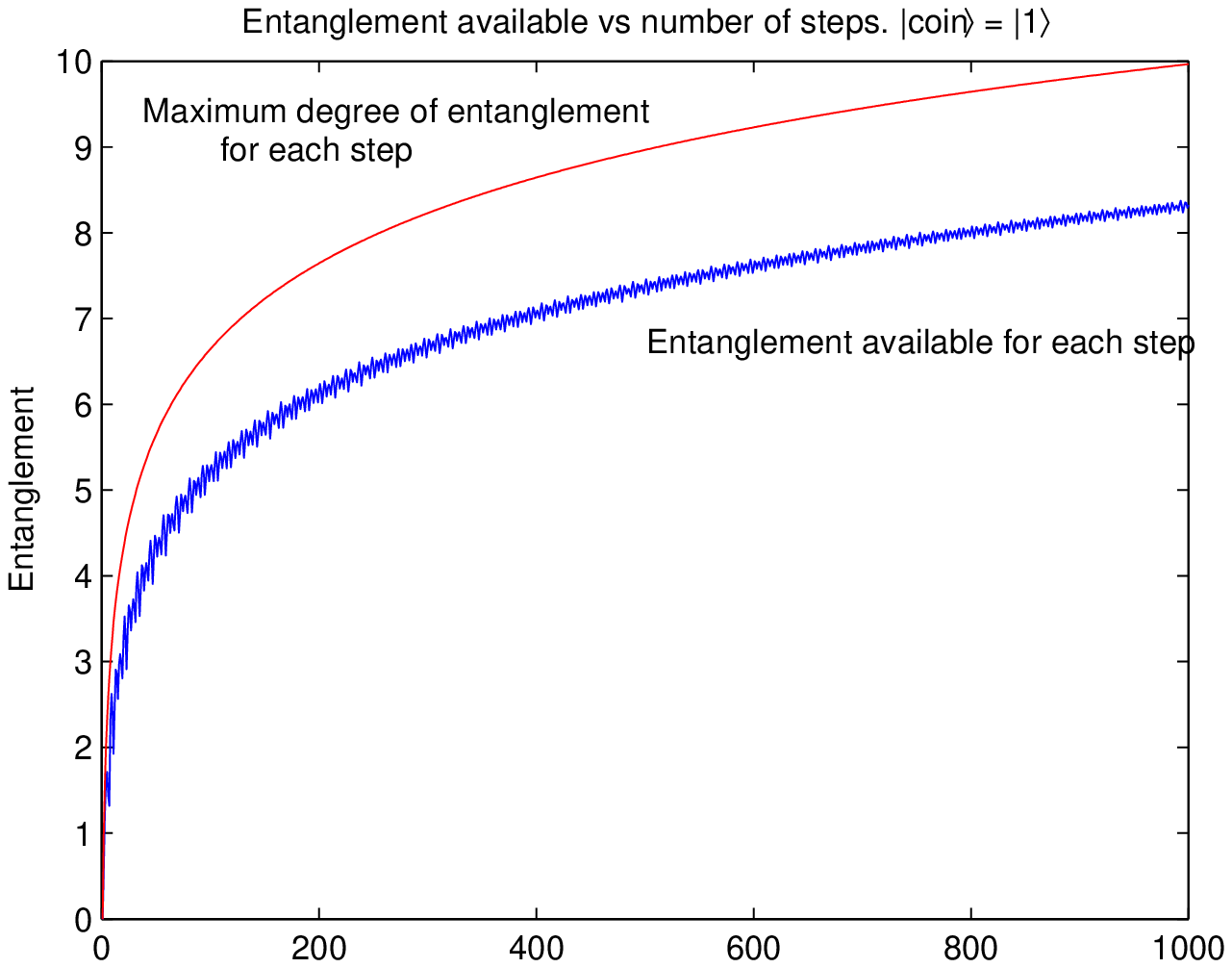}
(ii)\epsfig{width=1.5in,file=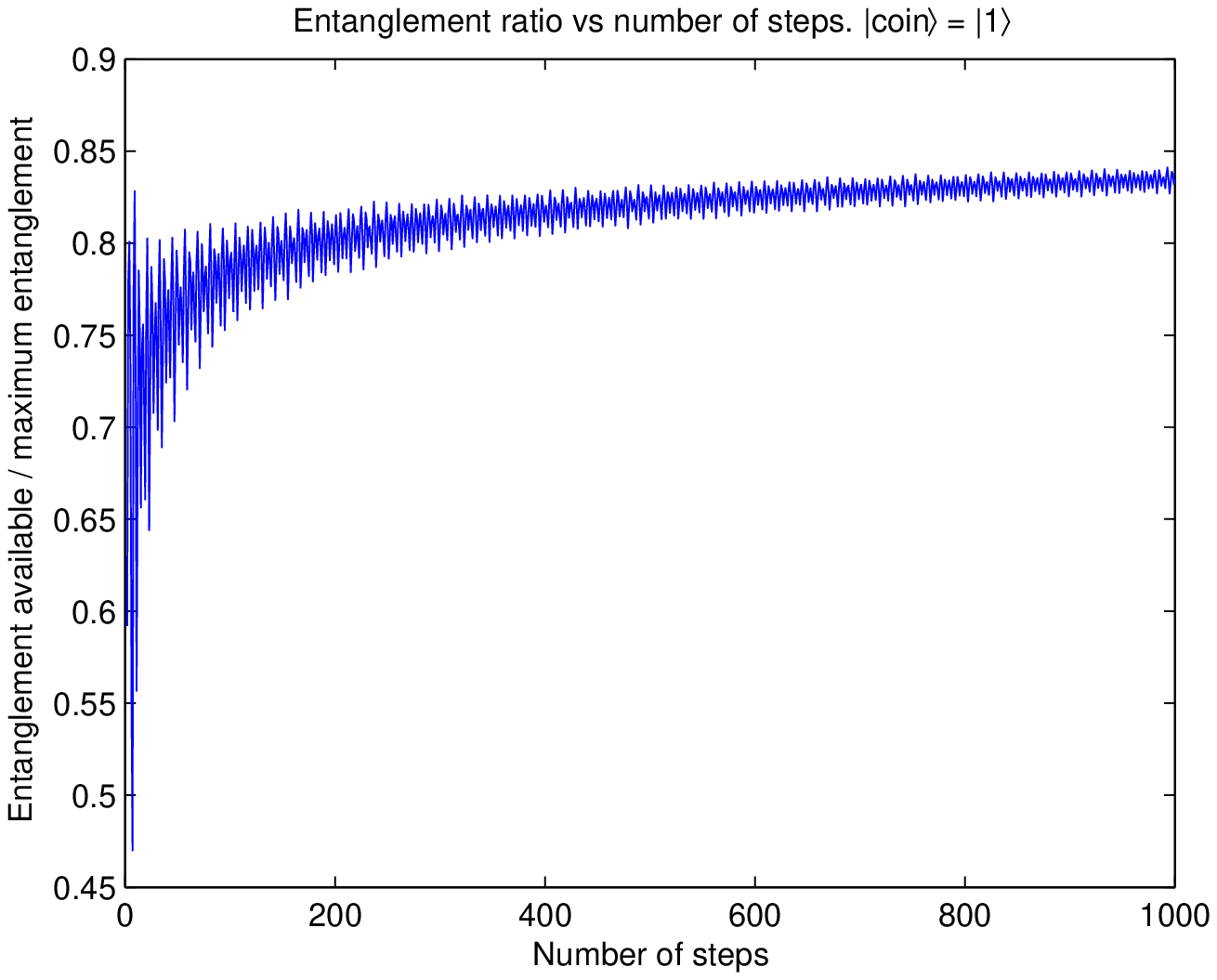}
\caption{{\small Entanglement values for coin $|1\rangle_c$ post-measurement state
$|\psi\rangle_{t,pm}^{c_1}$ computed from a 1000-steps quantum walk
$|\psi\rangle_{1000} = [{\hat S_{\text{ent}}} (\hat{H} \otimes {\hat I} )]^{1000} |\psi\rangle_0$
with $|\psi\rangle_0 = (\sqrt{0.85}|0\rangle_c - \sqrt{0.15}|1\rangle_c) \otimes |0,0\rangle_p$
given by Eq. (\ref{five_initial_condition}), coin (${\hat H}$) and shift (${\hat S}$) operators given by
Eqs. (\ref{hadamard_chapter_qwec_ii}) and (\ref{shift_operator_quant_entanglement})
respectively, and measurement operator ${\hat M}_1$ (Eq. (\ref{observable_qw})).
The thin line of (i) (red color online) shows the maximum degree of entanglement between
walkers attainable in the post-measurement quantum state $|\psi\rangle_{t,pm}^{c_1}$,
and the thick line of (i) (blue color online) shows the actual entanglement between walkers available
at each step. We can see that the asymptotics of entanglement values given in
plot (ii) tend to the same values as those shown in Fig. (\ref{fifth_condition_coin_zero}),
obtained from a coin $|0\rangle_c$ post-measurement state $|\psi\rangle_{t,pm}^{c_0}$
computed from a quantum walk with the same initial state (Eq. (\ref{five_initial_condition})).
}}
\label{fifth_condition_coin_one}
\end{figure}

\begin{figure}[hbt]
(i)\epsfig{width=1.5in,file=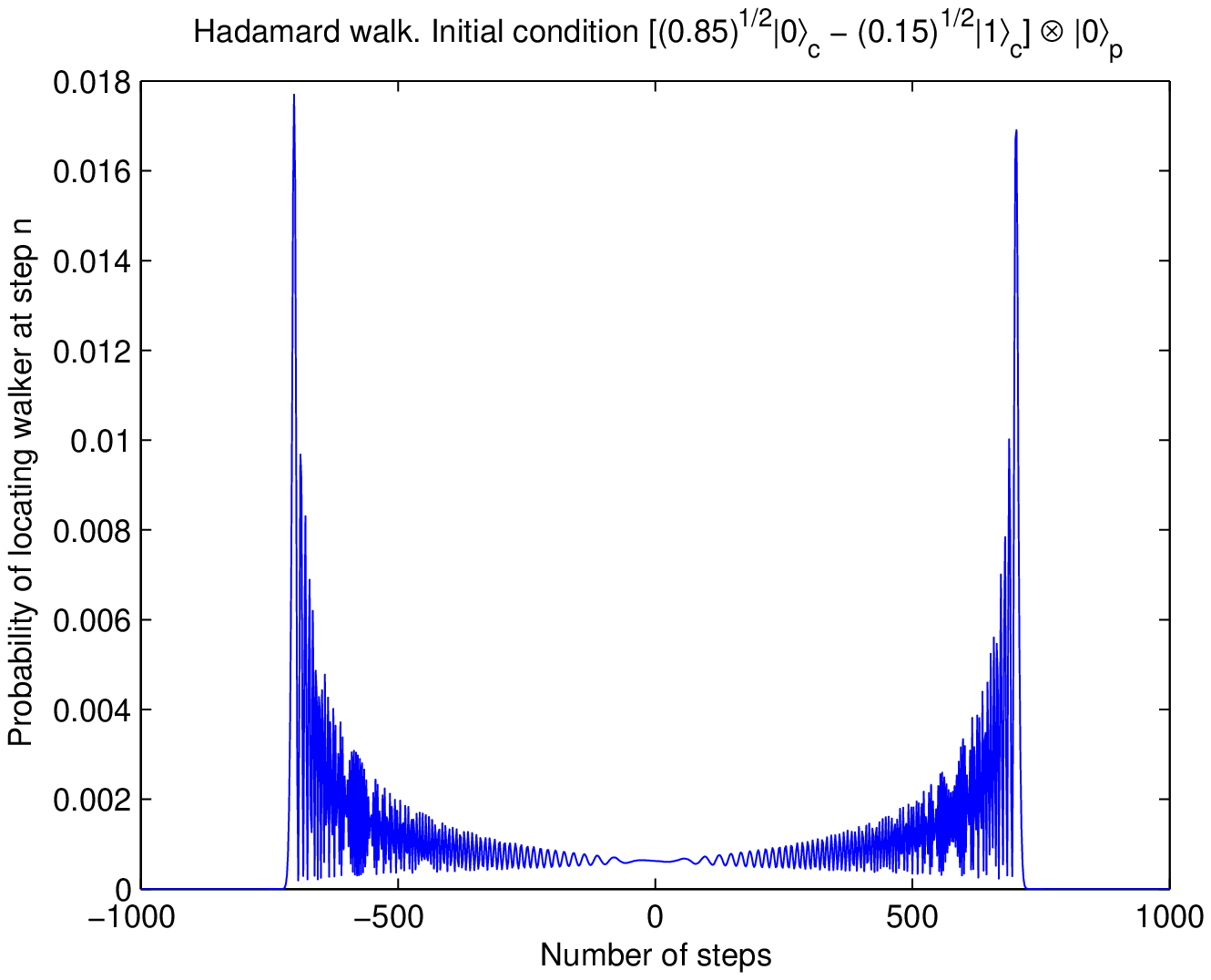}
(ii)\epsfig{width=1.5in,file=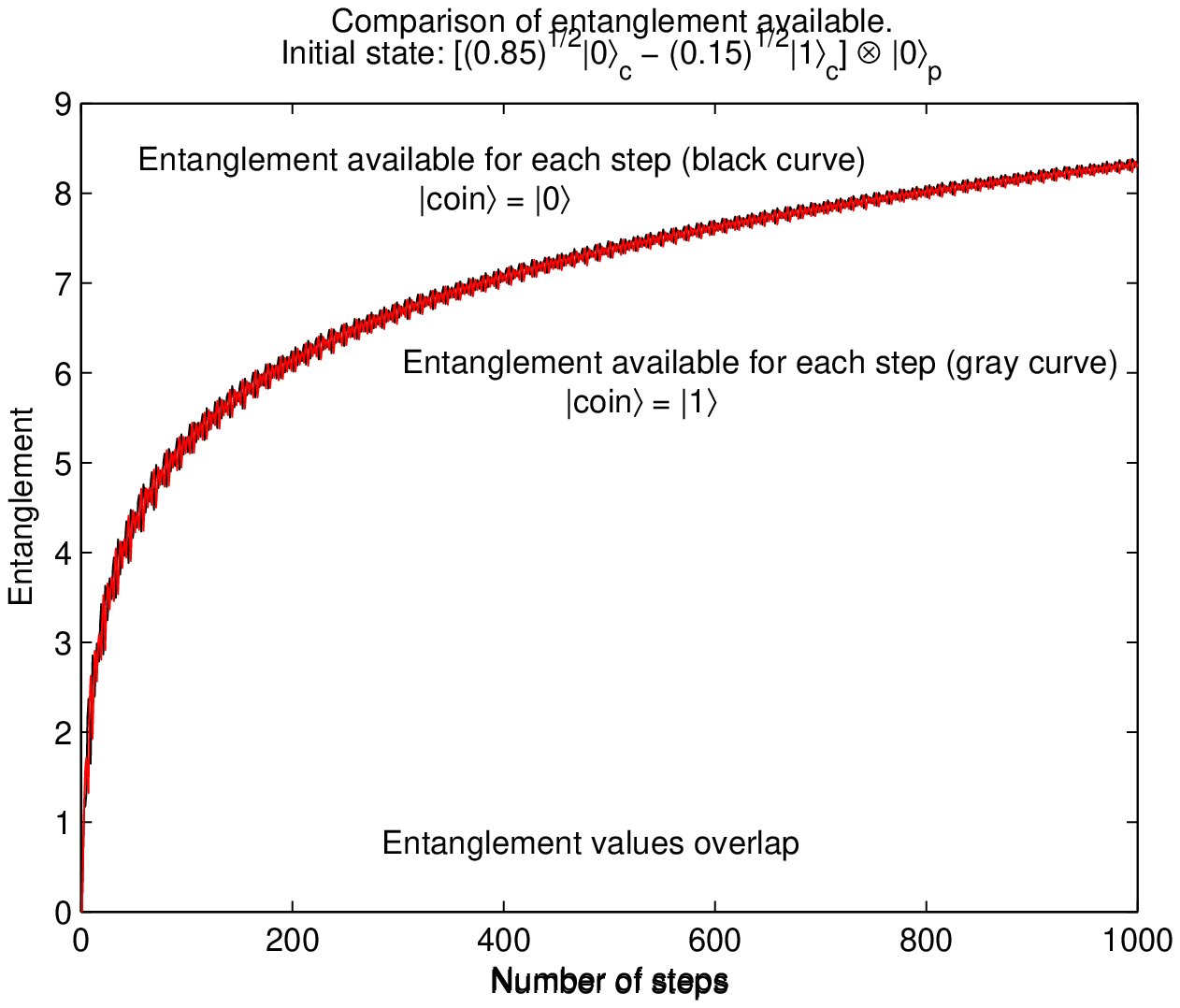}
\caption{{\small Plot (i) presents the probability vs location graph
of a 1000-step Hadamard quantum walk with an initial state
$(\sqrt{0.85}|0\rangle_c - \sqrt{0.15}|1\rangle_c) \otimes |0\rangle_p$
and shift operator provided by Eq. (\ref{shift_single}).
The symmetry of the probability distribution shown in plot (i) is the same
as that of a Hadamard quantum walk with initial state given by
$|\psi\rangle = (\sqrt{0.85}|0\rangle_c - \sqrt{0.15}|1\rangle_c) \otimes |0,0\rangle_p$ and
shift operator given by Eq. (\ref{shift_operator_quant_entanglement}).
In this case, the asymptotics of entanglement values for both coin post-measurement states
$|\psi\rangle_{t,pm}^{c_0}$ (black curve of plot (ii)) and $|\psi\rangle_{t,pm}^{c_1}$ (gray curve of plot (ii))
tend to the same values.
}}
\label{fifth_condition_entanglement_comparison}
\end{figure}

However and in stark contrast to the previous cases, the symmetry properties
of the probability distribution of a quantum walk with initial state
$|\psi\rangle_0 = (\sqrt{0.85}|0\rangle_c - \sqrt{0.15}|1\rangle_c) \otimes |0,0\rangle_p $
(Eq. (\ref{five_initial_condition})) does have an effect of the entanglement between walkers
produced from coin post-measurement quantum states.

In Figs. (\ref{fifth_condition_coin_zero}) and (\ref{fifth_condition_coin_one})
we exhibit the asymptotical behavior of entanglement values of coin post-measurement
states $|\psi\rangle_{t,pm}^{c_0}$ and $|\psi\rangle_{t,pm}^{c_1}$ respectively,
for a quantum walk with initial state given by Eq. (\ref{five_initial_condition}).
As opposed to previous cases in which asymptotical values of entanglement between
walkers were different for post-measurement states $|\psi\rangle_{t,pm}^{c_0}$ and $|\psi\rangle_{t,pm}^{c_1}$,
we can see in Figs. (\ref{fifth_condition_coin_zero}) and (\ref{fifth_condition_coin_one})
that the asymptotics of both entanglement curves tend to the same efficiency of 85\%  approximately.
This tendency can also be seen in Fig. (\ref{fifth_condition_entanglement_comparison}.ii)
where we show that both entanglement curves overlap.


\section{Conclusions}

We have proposed an algorithm to generate entanglement between
walkers, after measuring the coin state, for a Hadamard quantum
walk with one (2-dimensional) coin and two walkers. Our numerical
simulations show that, asymptotically, the amount of entanglement
generated between walkers does not reach the highest degree of
entanglement possible at each step (purely from the dimensionality
of the space explored by the walkers in a given step) for either
coin measurement outcome. Nevertheless, our simulations also show
that the entanglement ratio ($=$ entanglement generated/highest
value of entanglement possible, for each step) tends to converge
to a rather high value (for example, to $0.8$ or $0.9$), and the
actual convergence value seems to depend on the coin initial state
and on the coin measurement outcome.

Convergence of entanglement ratio leads to a most interesting
result: the actual value towards which the entanglement ratio
converges, for each coin measurement outcome, depends on the
symmetry of the coin initial state. However, the relationship is
not straightforward, as it is possible to find two coin initial
states ($|\psi\rangle_0 = {1 \over \sqrt{2}}|0\rangle + {i \over
\sqrt{2}}|1\rangle$ and $|\phi\rangle_0 =\sqrt{0.85}|0\rangle -
\sqrt{0.15}|1\rangle$) such that, although both produce balanced
probability distributions, only one coin initial state
($|\phi\rangle_0$) makes the asymptotical values of entanglement,
for both coin measurements, converge to the same value. Going to
two walkers and exploring their entanglement can thereby reveal
differences in two quantum walks which are not differentiated
easily in the usual case of a single walker.

A noteworthy feature of our algorithm is the high amount of the
entanglement generated between the walkers which grows with the
number of steps. Our scheme is particularly applicable in physical
realizations where the coin is a qubit (such as an atom or a
superconducting qubit) which interacts with a distinct physical
system (such as an electromagnetic field mode) acting as a walker
\cite{sanders03,agarwal05,sanders08}. Taking two such walkers,
each being a distinct system, one can probe how entanglement is
generated between them through our algorithm. As such walkers do
not naturally interact with each other, using the coin (qubit)
system is the only way to entangle them. A recent circuit QED
suggestion for the physical implementation of a quantum walk
\cite{sanders08} even estimate a walk of a significant number of
steps to be carried out within the decoherence times of the
relevant physical systems. Simply enhancing such schemes to two
walkers would enable observing the entanglement generation
mechanism that we have presented and analyzed in this paper.

\section{Acknowledgements}

S.V-A. gratefully acknowledges useful discussions with Dr J.L. Ball as well as 
the support of SNI, CONACyT and Tecnol\'{o}gico de Monterrey Campus Estado de M\'{e}xico.
S.B. acknowledges the support of the EPSRC, UK, the QIP IRC
(GR/S82176 /01), the Royal Society and the Wolfson Foundation.

\bibliographystyle{plain}

\end{document}